\documentclass[twocolumn,showpacs,superscriptaddress,amsmath,amssymb,prb]{revtex4-1}

\usepackage[dvips]{graphicx}
\usepackage{dcolumn}
\usepackage{bm}

\begin{document}

\title{Magnetic-field-induced nonlocal effects on the vortex interactions in twin-free YBa$_{2}$Cu$_{3}$O$_{7}$}


\author{J.\,S.\,White,$^{1-3}$ R.\,W.\,Heslop,$^2$ A.\,T.\,Holmes,$^2$ E.\,M.\,Forgan,$^{2}$ V.\,Hinkov,$^{4,5}$ N.\,Egetenmeyer,$^1$ J.\,L.\,Gavilano,$^1$ M.\,Laver,$^{1,6,7}$ C.\,D.\,Dewhurst,$^{8}$ R.\,Cubitt,$^{8}$ and A.\,Erb$^{9}$}
\affiliation{
Laboratory for Neutron Scattering, Paul Scherrer Institut, CH-5232 Villigen, Switzerland\\
$^2$School of Physics and Astronomy, University of Birmingham, Edgbaston, Birmingham, B15 2TT, UK\\
$^3$Laboratory for Quantum Magnetism, Ecole Polytechnique F\'{e}d\'{e}rale de Lausanne, CH-1015 Lausanne, Switzerland\\
$^4$Quantum Matter Institute, University of British Columbia, Vancouver, British Columbia V6T 1Z1, Canada\\
$^5$Max Planck Institut f\"{u}r Festk\"{o}rperforschung, D-70569 Stuttgart, Germany\\
$^6$Materials Research Division, Ris\o~DTU, Technical University of Denmark, DK-4000 Roskilde, Denmark\\
$^7$Nano-Science Center, Niels Bohr Institute, University of Copenhagen, DK-2100 K\o benhavn, Denmark\\
$^8$Institut Laue-Langevin, 6 rue Jules Horowitz, 38042 Grenoble, France\\
$^9$Walther Meissner Institut, BAdW, D-85748 Garching, Germany\\}
\date{\today}

\begin{abstract}

The vortex lattice (VL) in the high-$\kappa$ superconductor YBa$_{2}$Cu$_{3}$O$_{7}$, at 2~K and with the magnetic field parallel to the crystal \textbf{c}-axis, undergoes a sequence of transitions between different structures as a function of applied magnetic field. However, from structural studies alone, it is not possible to determine precisely the system anisotropy that governs the transitions between different structures. To address this question, here we report new small-angle neutron scattering measurements of both the VL structure at higher temperatures, and the field- and temperature-dependence of the VL form factor. Our measurements demonstrate how the influence of anisotropy on the VL, which in theory can be parameterized as nonlocal corrections, becomes progressively important with increasing magnetic field, and suppressed by increasing the temperature towards $T_{c}$. The data indicate that nonlocality due to different anisotropies play important roles in determining the VL properties.

\end{abstract}



\maketitle
\section{INTRODUCTION}
\label{sec:1Int}

In an applied magnetic field, type-II superconductors can be investigated directly via studies of the magnetic vortex lattice (VL). One of the most direct experimental probes of the VL is small-angle neutron scattering (SANS). Being a bulk sensitive technique, SANS continues to play a pivotal role in clarifying the details of the mixed state in various classes of superconductors.~\citep{Esk97,Pau98,Ris98,Yet99,Lev02,Gil02,Lav06,Bia08,Ino10,Li11,Fur11} Typically, in SANS studies it is reported how the VL structure and coordination vary as functions of both magnetic field and temperature ($T$). The results are often used to provide evidence as to the sources of anisotropy that might cause the VL structure to deviate from the perfectly hexagonal coordination predicted for an isotropic material.~\citep{Abr57} By increasing the applied magnetic field, these anisotropies become increasingly influential on the inter-vortex interaction, and can cause transitions between different VL structures.

The simplest theoretical models that can be used to predict the occurrence of VL structure \emph{transitions} are those based on local electrodynamic theory (`London' theory) extended with nonlocal correction terms.~\citep{Not02} These correction terms provide a direct coupling between a system anisotropy and the VL properties. The size of the correction terms scale according to $\kappa^{-2}$, where $\kappa$ is the Ginzburg-Landau parameter. Hence for high-$\kappa$ materials such as cuprates, nonlocal corrections are often neglected and local theory is deemed sufficient to understand observation. However, VL studies of reasonably high-$\kappa$ materials, such as borocarbides~\citep{Esk97,Pau98} and $s$-wave V$_{3}$Si,~\citep{Yet99} show that in these materials it is necessary to consider nonlocal effects associated with a Fermi surface anisotropy in order to understand the field-dependent VL structure.~\citep{Kog96,Kog97a,Kog97b,Suz10} Furthermore, in high-$\kappa$ materials that exhibit a $d$-wave gap anisotropy, nonlocal corrections cannot be neglected because the nonlocal length-scale $\xi_{0}$ is momentum-dependent $\xi_{0}(\textbf{k})\propto 1/ \Delta_{\textbf{k}}$ and hence will be divergent at the nodes. As a direct consequence of these `$d$-wave' nonlocal effects, unconventional VL structures are predicted.~\citep{Aff97,Fra97,Ami98} More sophisticated models can also account for the effects due to anisotropy suggested by the aforementioned nonlocal theories. Within quasiclassical Eilenberger theory, both the Fermi surface and superconducting gap anisotropies can be considered in equal measure when evaluating the VL free energy, and again unconventional VL structures are expected as functions of both field and $T$.~\citep{Eil68,Ich99,Nak02,Suz10} Therefore, while nonlocal effects are often neglected in high-$\kappa$ materials, there is substantial evidence to suggest that they play a decisive role in determining the VL properties.

Our subject high-$\kappa$ superconductor is YBa$_{2}$Cu$_{3}$O$_{7-\delta}$, the mixed state of which has been under scrutiny more than twenty years.~\citep{Gam87,Mag95,Fis97,Son97,Son99,Aus08,For90,Yet93a,Yet93b,Kei93,Kei94,For95,Joh99,Bro04,Sim04,Whi08,Whi09} SANS studies in particular have made important contributions towards understanding the mixed state in this material,~\citep{For90,Yet93a,Yet93b,Kei93,Kei94,For95,Joh99,Bro04,Sim04,Whi08,Whi09} with the motivation to understand the physics of both the VL and the pairing state within an anisotropic host material that has a nodal gap symmetry. YBa$_{2}$Cu$_{3}$O$_{7-\delta}$ exhibits an orthorhombic crystal symmetry and is composed of stacked two-dimensional CuO$_{2}$ plane layers, and one-dimensional CuO chain layers. For optimally- and over-doped compositions, the CuO chains display both long-range order along the crystal \textbf{b}-axis, and metallic behavior.~\citep{Bas05} As a consequence it is expected that, in addition to CuO$_{2}$ plane states, electronic states primarily associated with the chains are also superconducting below $T_{c}$. This picture is supported by reported values of the in-plane penetration depth ratio, $\gamma_{\lambda}~(=\lambda_{a}/\lambda_{b}\propto\sqrt{m_{a}^{\ast}/m_{b}^{\ast}})$ lying in the range 1.2-1.5.~\citep{Bas95,Gag97,Age00,Bro04,Aus08,Whi09} While the physical origin of the chain pairing interaction remains to be definitively explained, current evidence suggests that it may originate from a proximity effect coupling between chain and plane states.~\citep{Atk95,Atk08,Kon10} A further consequence of the crystal orthorhombicity is that the predominantly $d_{x^{2}-y^{2}}$ order-parameter \emph{must} contain an additional and finite $s$-wave admixture.~\cite{Tsu00} Evidence for this admixture is provided by phase sensitive,~\cite{Kir06} tunneling,~\cite{Smi05} and $\mu$SR~\cite{Kha07} studies.


The striking result of earlier SANS studies on fully-doped, albeit crystallographically twinned YBa$_{2}$Cu$_{3}$O$_{7}$, was the observation that with increasing magnetic field, the VL structure evolves \emph{continuously} from a low field hexagonal structure towards an almost square structure by 11~T.~\citep{Bro04,Whi08} The physical origin of this transition was suggested to be due to the increased importance of the $d$-wave order parameter anisotropy as vortex cores move closer together at high field.~\citep{Ich99,Nak02,Suz10} However, it was also acknowledged that such a transition may also be driven by nonlocal effects associated with a band structure anisotropy.~\citep{Bro04,Whi08} More recently, we reported a SANS study of the VL in a sample of \emph{twin-free}, and fully oxygenated YBa$_{2}$Cu$_{3}$O$_{7}$.~\citep{Whi09} In this sample the pinning effects on the VL due to twin boundaries, a characteristic feature of all previous SANS studies,~\citep{Yet93a,Yet93b,Kei93,Kei94,For95,Joh99,Bro04,Sim04,Whi08} are almost entirely suppressed. For the first time, this allowed a clear observation of the \emph{intrinsic} VL structure, and we reported an entirely new sequence of field-dependent VL structures and structure transitions at 2~K, with all of the structure phases separated by first-order transitions.~\citep{Whi09}

In this paper, we extend our previous SANS study of twin-free YBa$_{2}$Cu$_{3}$O$_{7}$ with new measurements of both the $T$-dependent VL structure, and the field- and $T$-dependence of the VL form factor. Measurements of the form factor have proved to be remarkably useful in characterizing the mixed state of other superconductors,~\citep{Cub07,DeB07,Bia08,Den09,Whi10,Fur11} and are often used in conjunction with simple theories in order to extract measures of the penetration depth and coherence length. The results reported here represent the first detailed investigation of the VL form factor in YBa$_{2}$Cu$_{3}$O$_{7}$.

\section{EXPERIMENTAL METHOD}
\label{sec:2Exp}
The SANS investigations reported in this study have been carried out using the SANS-I and SANS-II instruments at the Paul Scherrer Institut, Villigen, Switzerland, the D11 instrument at Institut Laue-Langevin, Grenoble, France, and the SANS-NG3 instrument at the NIST Center for Neutron Research, Gaithersburg, USA. For all experiments, neutrons of a mean wavelength between 6 and 10~\AA~were selected with a 10\% FWHM spread in wavelength. The neutrons were collimated over distances between 6-14~m before the sample position, and the diffracted neutrons were collected by a position-adjustable two-dimensional multidetector.

All of the SANS measurements were performed using the same twin-free sample of YBa$_{2}$Cu$_{3}$O$_{7}$ that was also used in Ref.~\onlinecite{Whi09}. A mosaic of six single crystals was assembled with a total mass of $\sim$30~mg. Each single crystal was grown from flux within home-made BaZrO$_{3}$ crucibles.~\citep{Erb96} The as-grown crystals were individually detwinned under uniaxial stress at elevated $T$,~\citep{Lin92,Hin07} and subsequently oxygenated under a high pressure oxygen atmosphere to achieve the overdoped O$_{7}$ phase.~\citep{Erb99} The zero-field $T_{\textrm{c}}$ of the largest of the six crystals was measured using a SQUID magnetometer. The mid-point of the transition was found to be 89.0(5)~K which is consistent with the sample being maximally doped.~\citep{Lia06} The sample mosaic was assembled on a 1~mm thick Al plate with the crystal \textbf{c}-faces flat and co-aligned about the \textbf{c}-axis. X-ray Laue measurements showed the mosaicity of the co-alignment to be $\lesssim$1.5$^{\circ}$. The detwinning ratio of the mounted sample was determined from single crystal neutron diffraction measurements of the (100) and (010) nuclear peak intensities. No signal attributable to a minority domain was detectable within the sensitivity of the measurement, limiting the minority domain population to $<1$~\%.



\begin{figure*}
\includegraphics[width=0.32\textwidth]{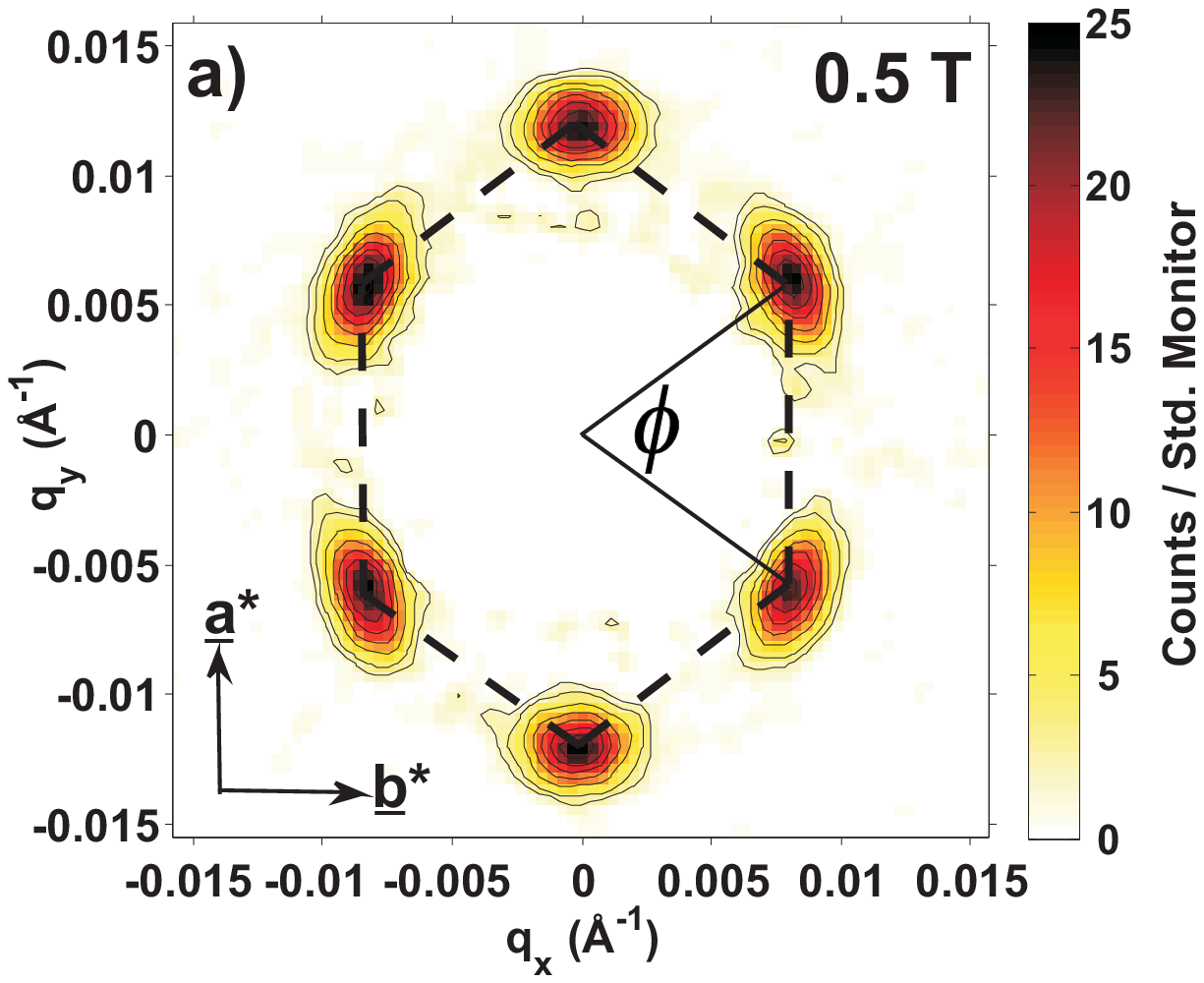}
\includegraphics[width=0.32\textwidth]{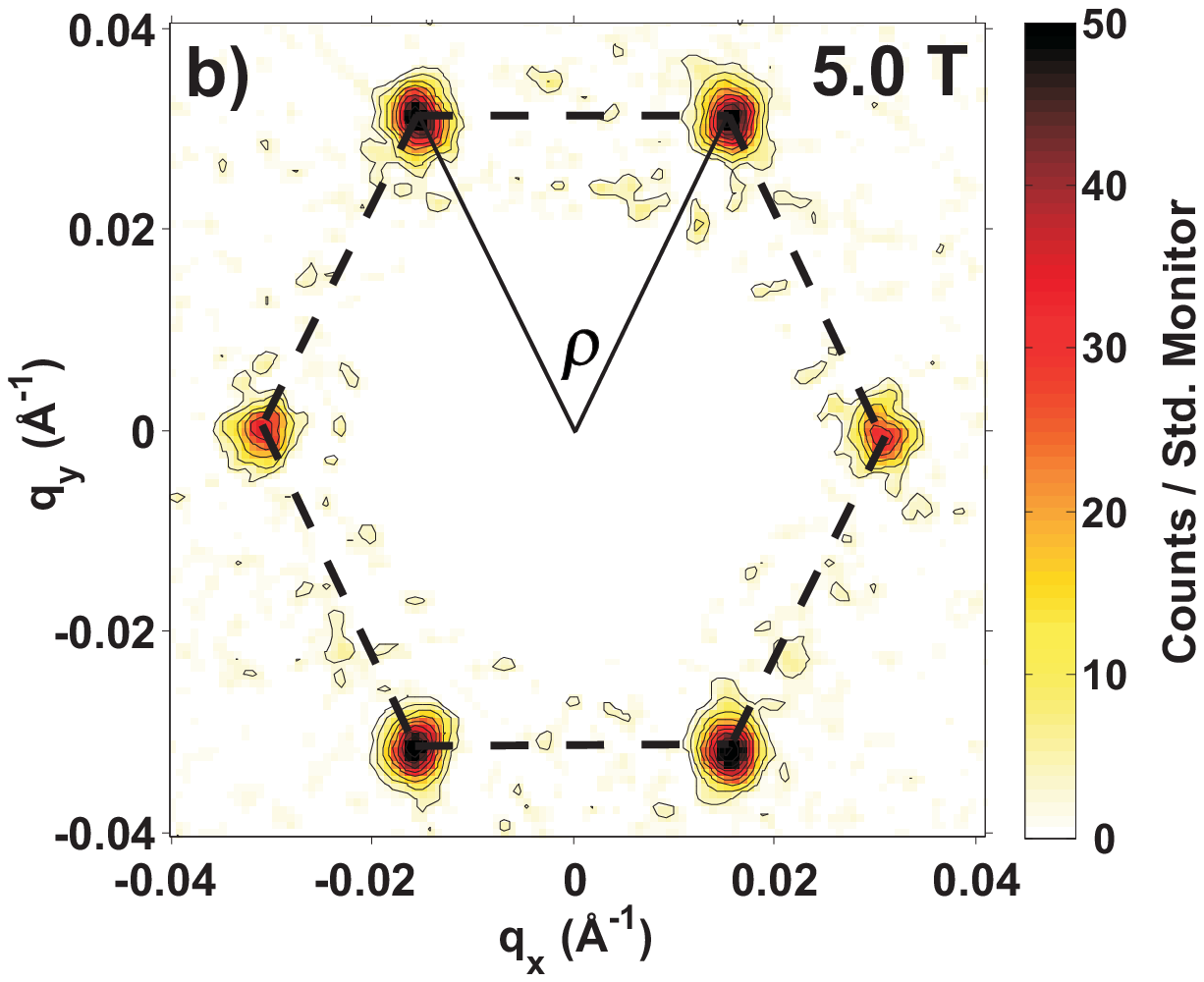}
\includegraphics[width=0.32\textwidth]{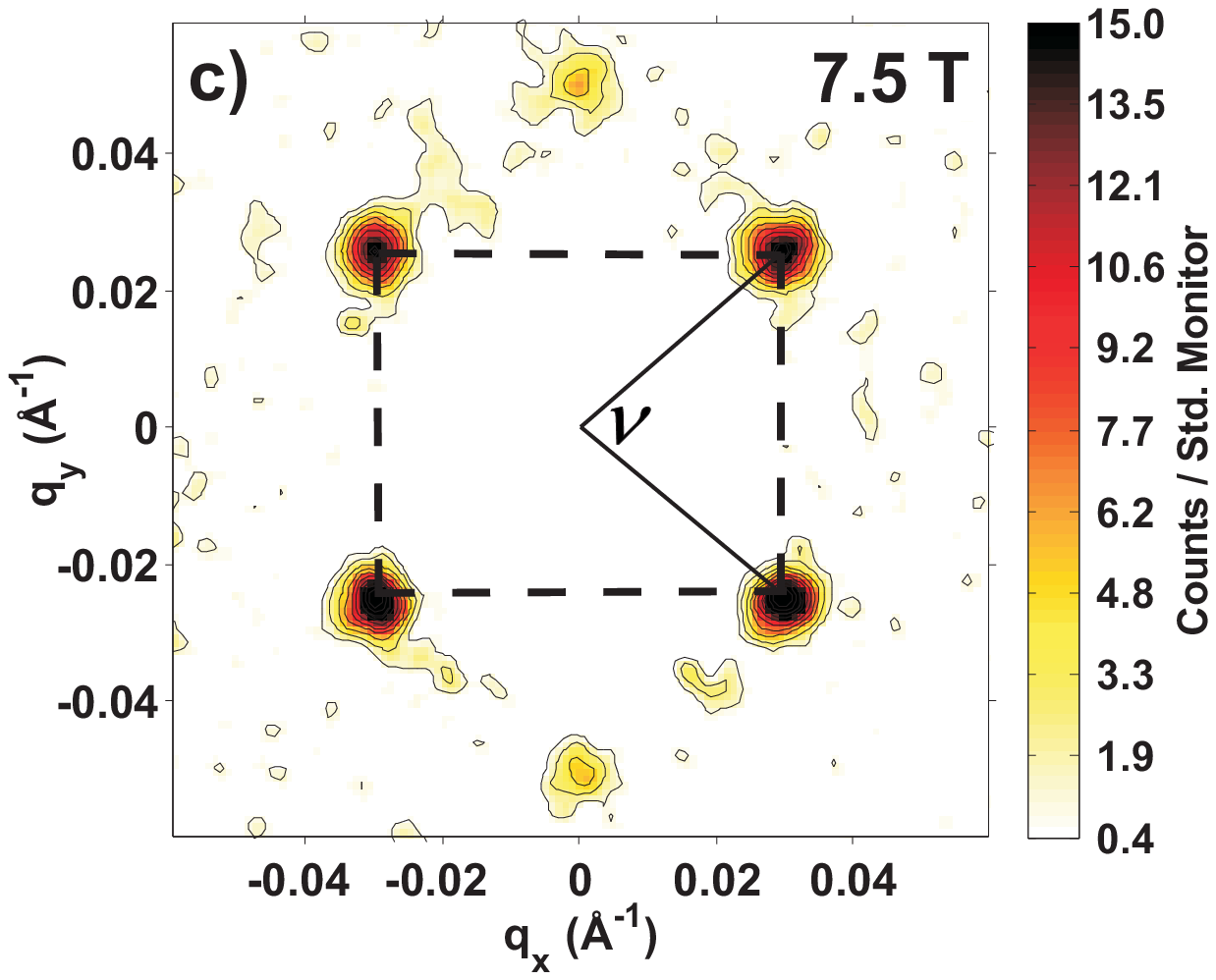}
\caption{(Color online) VL diffraction patterns obtained at 2~K, and in fields applied parallel to the crystal \textbf{c}-axis, of (a)~0.5~T, (b)~5.0~T and (c)~7.5~T. The axes indicated in (a) apply to all figures. In each image, dashed line patterns indicate the VL structures. Solid lines represent the basis vectors of the primitive cell, with the primitive cell opening angle of (a)~$\phi$, (b)~$\rho$ and (c)~$\nu$. The images are constructed by summing together the diffraction patterns obtained at a series of angles about the horizontal and vertical axes. This allows the presentation of all the Bragg spots in a single picture. Statistical noise at the center of the patterns has been masked, and the data smoothed with a Gaussian envelope smaller than the instrument resolution. The VL in real space can be visualized by rotating the reciprocal VL by 90$^{\circ}$ and adding an additional vortex at the origin.}
\label{fig:structures}
\end{figure*}

The sample mosaic was loaded into the variable-temperature-insert of a horizontal field cryomagnet with the \textbf{a}-axis vertical, and such that the \textbf{c}-axis was parallel to the field direction and also approximately parallel to the neutron beam. In our previous report,~\citep{Whi09} we described two different techniques for the preparation of the VL in the sample. The first is the standard field-cool (FC) procedure. In this case, the VL is prepared by applying the field above $T_{\textrm{c}}$, and then field-cooling through $T_{\textrm{c}}$ to the measurement $T$. The second is the oscillation field-cooling (OFC) procedure. In this case, the field-cool through $T_{\textrm{c}}$ is carried out in a field oscillating around the target value $B$, with an amplitude of typically 0.1~\% to 0.2~\% of $B$, and with an oscillation frequency of $\approx$~2~min$^{-1}$. We identified~\citep{Whi09} the role of the OFC procedure as that which equilibrates the VL structure into the preferred structural free energy minimum $\mathcal{F}$, that exists for an effective $T$ which is lower than the irreversibility temperature ($T_{\textrm{irr}}$) exhibited when using a standard field-cool. The OFC procedure is observed to be most effective at low fields, where the stabilization of the intrinsic VL structure is further accompanied by significant improvements in VL quality. For these reasons, in this paper we only report SANS measurements of the optimized VL prepared by the OFC procedure.

When at the desired measurement $T$, the OFC procedure is halted and the field is held constant at the target value. The subsequent diffraction measurements are carried out by rotating the sample and cryomagnet together about the horizontal and vertical axes to angles that satisfy the Bragg condition for a certain Bragg spot on the detector. In all cases, background measurements were taken at $T>T_{\textrm{c}}$, and subtracted from low $T$ foregrounds to leave just the VL signal. Data visualization and analysis was performed using the GRASP analysis package.~\citep{Dew03}

\section{RESULTS AND DISCUSSION}
\label{sec:3Res}

\subsection{Vortex Lattice Structure at 2~K}
\subsubsection{Field-Dependence at 2~K}
\label{sec:3Res_VL_Struct_2K}

To set the stage for the new results in this paper, we begin by reviewing our previous study.~\citep{Whi09} At 2~K, and within the field range up to 10.8~T, we observe a sequence of first-order VL structure transitions between the two orientations of the VL primitive cell permitted by the orthorhombic crystal symmetry.~\citep{Wal95} Both of these orientations as shown in the new SANS diffraction patterns presented in Fig.~\ref{fig:structures}. Fig.~\ref{fig:structures}~(a) shows a diffraction pattern obtained in an applied field of 0.5~T. The overlaid hexagonal pattern indicates that the VL structure is composed of a single distorted hexagonal domain aligned with the underlying atomic lattice. We refer to this structure type as the low field structure (LFS). The shape of the VL unit cell may be described by the angle $\phi$ which lies between the two basis vectors indicated in Fig.~\ref{fig:structures}~(a). On increasing the field, the VL structure undergoes a first-order 90$^{\circ}$ re-orientation transition into another single domain distorted hexagonal structure. Fig.~\ref{fig:structures}~(b) shows an example of this structure type in a diffraction pattern obtained at 5.0~T. We refer to this structure type as the intermediate field structure (IFS). The sign of the distortion of the IFS remains consistent with that of the LFS; both structures are stretched from an isotropic hexagonal coordination along the \textbf{a}$^{\ast}$ direction. The overlaid pattern in Fig.~\ref{fig:structures}~(b) indicates how we describe the primitive cell of the VL with a different characteristic angle $\rho$. On further increase of the field, we observe another first-order transition in the VL structure, this time between the IFS and a high field structure (HFS) that is dominated by a rhombic shape composed of four intense Bragg spots. This rhombic shape is highlighted in Fig.~\ref{fig:structures}~(c) which shows an example diffraction pattern obtained from the HFS in an applied field of 7.5~T. Since the primitive cell of the HFS is of the same orientation as that for the LFS, in order to distinguish between them we use the symbol $\nu$ to label the opening angle of the HFS primitive cell.

An interesting detail of the diffraction pattern shown in Fig.~\ref{fig:structures}~(b), and characteristic of the IFS phase, is that the horizontal spots with \textbf{q} parallel to the \textbf{b}$^{\ast}$ axis ($\textbf{q}\parallel \textbf{b}^{\ast}$) are weaker in intensity than the other four spots. This is converse to the general expectation that weaker spots appear at larger \textbf{q} than more intense spots and are consequently considered to be higher-order spots of the reciprocal VL. In the HFS phase, the more usual situation is realized; the weaker spots with \textbf{q} parallel to the \textbf{a}$^{\ast}$ axis ($\textbf{q}\parallel \textbf{a}^{\ast}$) exhibit a larger \textbf{q} than the four strong spots that make up the rhombic structure, and thus may be considered as higher-order spots. According to this description, we see that the rhombic shape is distorted away from an isotropic square lattice by a stretching along the \textbf{b}$^{\ast}$ direction. Such a distortion is of opposite sign to that observed for the hexagonal structures; a possible physical reason for this is discussed later. Alternatively, the HFS may also be considered as a distorted hexagonal structure of the same orientation as the LFS phase, and the same direction of distortion as the LFS and IFS phases. While there is no formal method to distinguish between such descriptions of the VL structure, other data presented later will show that with increasing field, the spots of the rhombic structure become further dominant in intensity over the weak spots with $\textbf{q}\parallel \textbf{a}^{\ast}$.

\begin{figure}
\includegraphics[width=0.5\textwidth]{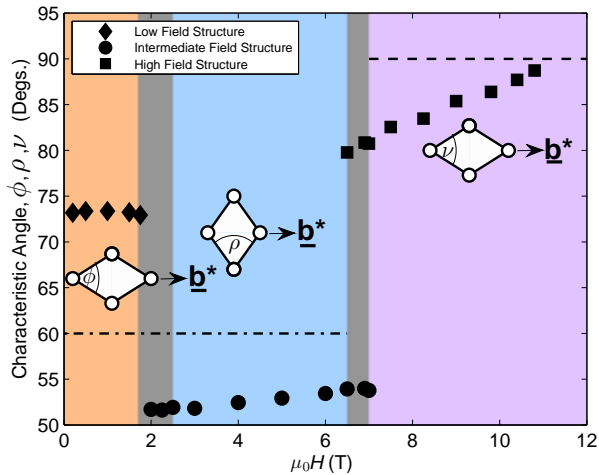}
\caption{(Color online). The field-dependence at 2~K of the primitive cell opening angle for the various VL structures. Inset schematics indicate the reciprocal VL primitive cell for each phase. The dashed line at 90$^{\circ}$ indicates the opening angle of a perfectly square VL. The dashed-dot line at 60$^{\circ}$ indicates the opening angle of an isotropic hexagonal VL. The (dark) gray shaded portions indicate the field regions over which we observe first-order VL structure transitions. The error bars are sufficiently small to be masked by the data symbols. Part of these data were first presented in Ref.~\onlinecite{Whi09}.}
\label{fig:opening_angles}
\end{figure}
In Fig.~\ref{fig:opening_angles}, we show a quantitative analysis of the VL structure and plot the field-dependence of the primitive cell opening angle at 2~K. At 2.1(4)~T and 6.7(2)~T we observe clear discontinuities in the overall field-dependence, confirming the first-order nature of the VL transitions. The widths in field of the regions of structure co-existence are estimated to be 0.75(5)~T for the low field transition, and 0.45(5)~T for the high field transition. The finite width of these co-existence regions may arise from slight differences between the various crystals of the sample mosaic or from hysteresis within them, though the broader LFS - IFS transition also indicates that the LFS and IFS types are closer in free energy than the IFS and HFS phases. Fig.~\ref{fig:opening_angles} also shows that the primitive cell angle of the HFS is still evolving with field until the highest fields available. Whilst at no field do we observe a perfectly square coordination with a primitive cell angle of 90$^{\circ}$, we stress that in orthorhombic YBa$_{2}$Cu$_{3}$O$_{7}$ there is no particular symmetry reason for the VL to stabilize into a perfectly square structure at high fields.


\begin{figure}
\includegraphics[width=0.5\textwidth]{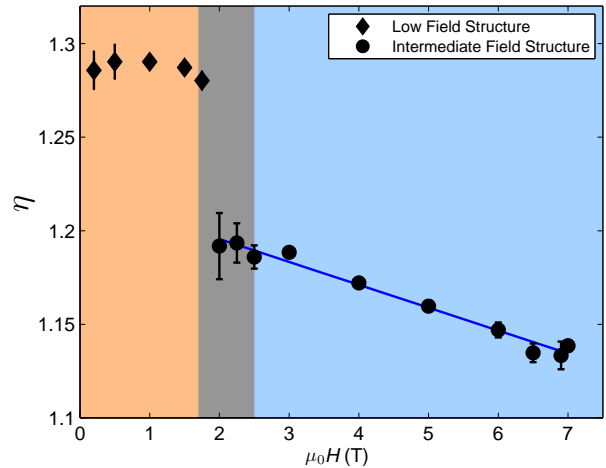}
\caption{(Color online). The field-dependence at 2~K of the parameter $\eta$ determined for both the LFS and IFS. The (dark) gray shaded portion indicates the field region over which we observe a first-order transition between the LFS and IFS phases. The line corresponds to a linear fit of the data obtained in the IFS phase. Error bars not visible can be considered of order the size of the symbol. Part of these data were first presented in Ref.~\onlinecite{Whi09}.}
\label{fig:hex_distortion}
\end{figure}


Next we examine the distortion of the VL structure in the LFS and IFS phases. The stretching of the diffraction patterns along \textbf{a}$^{\ast}$ is consistent with that expected due to chain enhanced superconductivity along the \textbf{b} direction. This anisotropy is incorporated into local London theory by inserting a second-rank effective mass tensor into the expression for the London free energy.~\citep{Kog81,Cam88,Thi89} For the uniaxial case where $m_{a}^{\ast}=m_{b}^{\ast}$, $\gamma_{\lambda}=\sqrt{m_{a}^{\ast}/m_{b}^{\ast}}=1$, the VL structure is undistorted and perfectly hexagonal. In such a case, the spot distribution can be overlaid by a circle. For the bi-axial case $m_{a}^{\ast}\neq m_{b}^{\ast}$, the VL structure in reciprocal space is scaled by $\gamma_{\lambda}^{1/2}$ along \textbf{a}$^{\ast}$ and by $\gamma_{\lambda}^{-1/2}$ along \textbf{b}$^{\ast}$, resulting in an elliptical spot distribution.~\citep{Cam88,Thi89} In the anisotropic London limit, the axial ratio of the ellipse, which we call $\eta$, gives directly the value of the anisotropy parameter $\gamma_{\lambda}$. Geometry shows that for the LFS, $\eta$ can be determined directly from the defined opening angle $\phi$ according to $\eta_{LFS}=\sqrt{3}/\textrm{tan}\left(90^{\circ}-(\phi/2)\right)$. For the IFS, $\eta$ can be found from the angle $\rho$ according to $\eta_{IFS}=(\sqrt{3}\textrm{tan}\left(\rho/2\right))^{-1}$. Fig.~\ref{fig:hex_distortion} show the field-dependence at 2~K of $\eta$ obtained within both the LFS and IFS phases. For the LFS phase, the value of $\eta$ remains essentially field-independent. The mean value taken across this phase is $\eta=$~1.28(1), a value similar to those found in previous studies,~\citep{Bas95,Age00,Bro04,Aus08} some of which were performed on YBa$_{2}$Cu$_{3}$O$_{7-\delta}$ samples of slightly different composition. On moving between the LFS and IFS phases, the first-order change in the VL structure is accompanied by a discontinuity in $\eta$. Within local London theory, this would imply a similarly sharp discontinuity in the value of $\gamma_{\lambda}$ which is not expected. This suggests that nonlocal effects are becoming important. On further increase of the field, the value of $\eta$ is observed to fall monotonically, which similarly within the local theory would indicate a suppression of the in-plane anisotropy. However, results in the following sections will show that such a simple picture may be inappropriate, and that a more detailed theory is required in order to explain the observations.


\subsubsection{Discussion}
Within the literature, a broad range of theoretical work attempts to describe the field-dependent behavior of the low $T$ VL structure in high-$\kappa$ materials. In our case, the simplest model is the anisotropic London theory. This is a purely local theory that is valid for $\kappa\gg1$ and for fields $B\ll B_{\textrm{c2}}$, a regime deemed suitable for experiments on High-$T_{\textrm{c}}$ materials.~\citep{Kog81,Cam88,Thi89} According to local theory, the preferred VL structure is expected to be hexagonal in coordination, but with no preferred structural \emph{orientation}. Since this orientational degeneracy is expected within local theory, even an extremely weak interaction may lead to a preferred orientation. Therefore local theory may still be dominant at low fields, even though preferred orientations are seen. Since no field-driven VL structure transitions are expected from the predictions of local theory, this demonstrates that the local approach neglects important details, even if the theory is able to describe the VL distortion in high-$\kappa$ materials.

An orientational $\mathcal{F}$ can be found at every field and $T$ after extending the local London theory with nonlocal corrections that take the form of higher-order terms in the expression for the London free energy. These higher-order terms provide a coupling of the VL both to the anisotropy of the band structure~\citep{Kog96,Kog97a,Kog97b,Fra97,Suz10} and, in general, an anisotropy of the gap.~\citep{Aff97,Fra97,Ami98} In each case, numerical evaluations of the free energy are able to reproduce field-driven and first-order re-orientation transitions between hexagonal VL structures that are analogous to our observed LFS - IFS transition. However, nearly all model predictions are made using an idealized fourfold symmetric system, and predictions for twofold symmetric systems are comparatively few.~\citep{Kog97a,Kog97b,Hir10} Seemingly the most suitable predictions for twofold symmetric systems such as YBa$_{2}$Cu$_{3}$O$_{7}$ are provided by the nonlocal theory of Kogan~\emph{et al.}~\cite{Kog97a,Kog97b} However, before we discuss the application of the theory to our results,~\cite{Kog97a,Kog97b} we next discuss the theory in more detail, because we have discovered it to possess some previously unreported properties.

Kogan's nonlocal theory~\cite{Kog97a} accounts for the effects of Fermi-surface anisotropy, but not gap anisotropy, on the VL structure (although we expect that a $d$-wave gap anisotropy will have similar effects~\cite{Fra97}). The theory is formulated by developing the lowest-order nonlocal corrections to the local London relationship between supercurrent density, $\textbf{j}$ and magnetic vector potential, $\textbf{a}$. Kogan~\emph{et al.} showed that by including these corrections the general form of the nonlocal London equation becomes~\cite{Kog96,Kog97a,Kog97b}
\begin{eqnarray}\label{Kog1}
\mu_{0}j_{i}=-\lambda^{-2}\left(m_{ij}^{-1}-\lambda^{2}n_{ijlm}G_{l}G_{m}\right)a_{j},
\end{eqnarray}
where, $m_{ij}$ is a normalized effective mass tensor and $\lambda$ is the geometrically averaged penetration depth $=(\lambda_{x}\lambda_{y}\lambda_{z})^{-1/3}$. Nonlocality is captured by the term dependent on both the reciprocal VL vector \textbf{G}, and the fourth tank tensor, $n_{ijlm}$. This tensor provides a coupling between the VL supercurrents and the microscopic properties of the Fermi surface, and is dependent on the fourth moment of the Fermi velocity according to $n_{ijlm}\propto\langle v_{i}v_{j}v_{l}v_{m}\rangle$.~\citep{Kog96,Kog97a,Kog97b} Without the term dependent on $n_{ijlm}$, Eq.~\ref{Kog1} is the local anisotropic London equation.



The free energy density, $\widetilde{F}$ of a certain VL coordination is computed numerically by summing the contributions due to the various Fourier components of the field distribution described by Eq.~\ref{Kog1}. At each field, the preferred VL coordination is that which minimizes $\widetilde{F}$. For fields applied along a fourfold symmetric $z$-axis, the $x$ and $y$ directions are equivalent, and $\widetilde{F}$ is evaluated by performing the sum:~\cite{Kog97a}
\begin{gather}
\widetilde{F}=\frac{B^{2}}{2\mu_{0}}\sum_{\textbf{G}}\frac{1}{\left(1+\lambda^{2}G^{2}+\lambda^{4}\left(n_{2}G^{4}+dG_{x}^{2}G_{y}^{2}\right)\right)}\nonumber \\
d=2n_{1}-6n_{2}\label{Kog2}
\end{gather}
where $n_{1}\propto \langle v_{x}^{4}\rangle \equiv \langle v_{y}^{4}\rangle$ and $n_{2}\propto\langle v_{x}^{2}v_{y}^{2}\rangle$. When evaluating $\widetilde{F}$, the logarithmic infinity in the magnetic field at a vortex core given by London theory has to be regularized. This is done by inserting a `core-correction' term of the form $\textrm{exp}\left(-G^{2}\xi^{2}\right)$ into the numerator of Eq.~\ref{Kog2}, and may itself be anisotropic due to effective mass anisotropy. Since effective mass is represented by a second rank tensor, for the fourfold case, the effective mass of the carriers is isotropic in the basal plane and the effects of the core corrections~\cite{Kog97a,Kog97b} are thus taken to be isotropic too. Similarly the local limit for the value of the penetration depth is also isotropic since it too depends on the effective mass. Therefore, the nonlocal effects are the only cause of anisotropy.


For fields applied along a twofold symmetric $z$-axis, (appropriate for the case of $\mu_{0}H \parallel \textbf{c}$ in YBa$_{2}$Cu$_{3}$O$_{7}$) the effective mass is anisotropic (the $x$ and $y$ directions are inequivalent) so both the value of penetration depth and the core correction are modified to take this into account.~\cite{Kog97a} Kogan~\emph{et al.} showed that the general expression for $\widetilde{F}$ becomes:~\cite{Kog97a}
\begin{gather}
\widetilde{F}=\frac{B^{2}}{2\mu_{0}}\sum_{\textbf{G}}\frac{1}{\left(1+\lambda^{2}K^{2}+\lambda^{4}\left(n_{4}K^{4}+bG_{x}^{2}G_{y}^{2}\right)\right)}\nonumber \\
K^{2}=(m_{x}G_{x}^{2}+m_{y}G_{y}^{2}),\nonumber \\  b=n_{3}m_{y}^{2}+n_{1}m_{x}^{2}-6n_{4}m_{x}m_{y}\label{Kog3}
\end{gather}
where $m_{i}$ is the effective mass in the $i$th direction, and the tensors $n_{3}\propto \langle v_{x}^{4}\rangle$, $n_{1}\propto \langle v_{y}^{4}\rangle$ and $n_{4}\propto\langle v_{x}^{2}v_{y}^{2}\rangle$.

\begin{figure}
\includegraphics[width=0.40\textwidth]{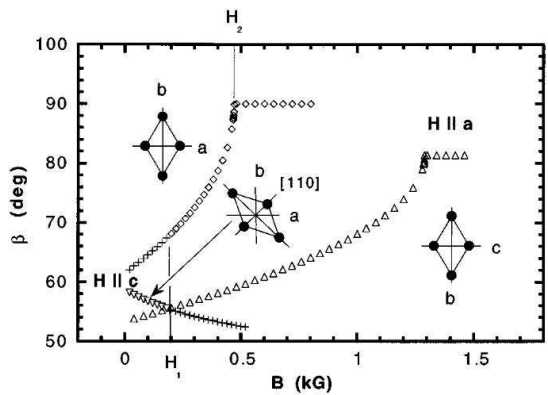}
\caption{The predicted field-dependences of the VL opening angle $\beta$ for both the fourfold symmetric case (down pointing triangles and diamonds), and the twofold symmetric case (up pointing triangles). Figure taken from Ref~\onlinecite{Kog97a}.}
\label{fig:kogan_prediction}
\end{figure}

By using Eqs.~\ref{Kog2}~and~\ref{Kog3}, Kogan~\emph{et al.} calculated the field evolution of the VL structures for both the fourfold and twofold symmetric cases.~\citep{Kog97a} Their reported predictions are shown in Fig.~\ref{fig:kogan_prediction}. For the fourfold case, the hexagonal VLs predicted in the low- and mid-field ranges break the fourfold crystal symmetry. Hence at any field where hexagonal VLs are expected, there are two degenerate VLs rotated 90$^{\circ}$ to one other. The lowest-field hexagonal VLs are predicted to be oriented at 45$^{\circ}$ to the mid-field hexagonal VLs, and change from one to the other via a first-order transition. The mid-field VLs then distort with increasing field before locking via a second-order transition into a single square VL of definite orientation relative to the crystal axes.

For the twofold symmetric case no low-field hexagonal VL expected, and the analogue of the mid-field hexagonal VL is present in only one orientation, aligned with one of the crystal axes. This VL is predicted to distort as the field is increased and to lock-in via a second-order transition to a non-square high-field VL with field-independent shape, and again in a definite orientation relative to the crystal axes. Whereas the fourfold symmetric version of the theory gives reasonable agreement with observations in superconductors with appropriate field directions,~\cite{Kog97b,Pau98,Yet99} it does not agree with those obtained with the field along a twofold symmetric axis.~\cite{Esk01}

We have therefore reconsidered the twofold symmetric version of the theory, and discovered that it has some previously unreported properties. Starting in the twofold symmetric case, we can make a scale transformation of the $x$- and $y$-components of the spatial variation of the magnetic field: $m_{x} (G_{x})^{2} \rightarrow (G_{x}')^{2}$ and $m_{y} (G_{y})^{2}  \rightarrow  (G_{y}')^{2}$, and thus remove the effective mass anisotropy from the theory. Now all the local and nonlocal contributions to the penetration depth become identical in form to the fourfold symmetric case (with coefficients derived from the twofold parameters also rescaled). This \textbf{k}-space scaling is also exactly that required to make the core corrections in the theory isotropic. Hence, the twofold symmetric version of the theory has \emph{exactly} the same structure as the fourfold symmetric version, except that the scale transformation distorts the shape of the predicted VLs and re-scales the value of the field.

\begin{figure}
\includegraphics[width=0.45\textwidth]{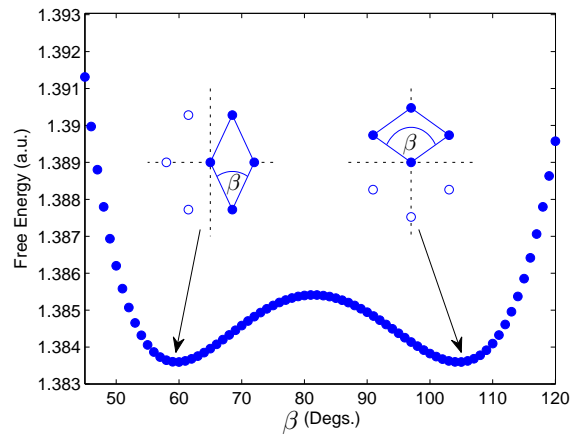}
\caption{Numerical evaluations of the free energy carried out using the Kogan nonlocal theory in the mid-field region and for the twofold symmetric case.~\citep{Kog97a} Inset diagrams show the degenerate reciprocal VLs. Filled symbols denote the primitive cell, and empty symbols represent the additional spots required to complete the hexagonal structure. Dashed lines indicate horizontal and vertical symmetry planes. The parameters used for the calculations are the same those in Ref.~\onlinecite{Kog97a}.}
\label{fig:kogan_one}
\end{figure}

\begin{figure}[t]
\includegraphics[width=0.45\textwidth]{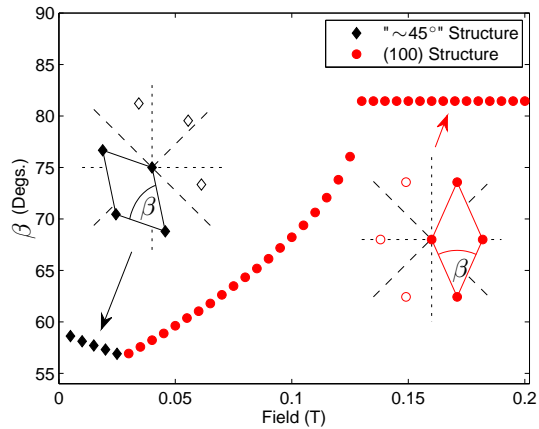}
\caption{The field-dependence of the primitive cell opening angle $\beta$ as calculated using the Kogan theory for the twofold symmetric case. In the mid-field region, we plot only the small-$\beta$ solution in order to compare with Fig.~\ref{fig:kogan_prediction}. The reciprocal VL structures are defined by the inset diagrams. The short- and long-dashed lines indicate the crystal symmetry planes for the \emph{fourfold} case. For the \emph{twofold} case, only the short-dashed line symmetry planes remain. The parameters used for the calculations are the same as those in Ref.~\onlinecite{Kog97a}.}
\label{fig:kogan_two}
\end{figure}

Using Eq.~\ref{Kog3}, we have made a numerical implementation of the theory for the twofold case in order to study the consequences of the isomorphism of the twofold and fourfold theories. Firstly, we find that in the mid-field region, two degenerate hexagonal VLs are expected [Fig.~\ref{fig:kogan_one}], instead of just the single hexagonal VL (at low $\beta$) that was identified previously.~\citep{Kog97a} This degeneracy may be expected for the twofold case since for the fourfold symmetric case, two identical (but rotated by 90$^{\circ}$) distorted hexagonal VLs have the same free energy. Secondly, we find that the minimum free energy at the \emph{lowest} fields corresponds to VLs in orientations that are rotated approximately 45$^{\circ}$ from the mid-field VLs as shown in Fig~\ref{fig:kogan_two}.~\citep{Not03} After scaling to go from the fourfold to the twofold symmetric case, these lowest field VLs are in unsymmetrical orientations relative to the crystal axes, and so were not searched for in the earlier work. Again, in agreement with the fourfold symmetric case, the low- and mid-field VL structures are separated by a first-order reorientation transition of the primitive cell. In the high-field region, the expected single non-square VL solution corresponds to a fourfold square VL distorted by the scaling factors.

We also observe that there is no change of symmetry expected at the transition between the mid-field hexagonal VL and the high-field non-square VL [Figs.~\ref{fig:kogan_prediction},~\ref{fig:kogan_two}]. This is of concern since in a \emph{real} twofold symmetric system there \emph{cannot} be a second-order transition between such VLs: it is only predicted in the twofold case because of its `hidden fourfold symmetry'. This symmetry is a property of the theory and \emph{not} a necessary property of the system being described. Hence, in practice, any property of the system not contained within the theory will remove the sharp transition in the twofold case, and replace it with a smooth crossover. We can remove this hidden symmetry easily by putting a different effective mass anisotropy into the core correction from that in the penetration depth. This maintains the twofold crystal symmetry, but removes the isomorphism of the two- and a fourfold theories. Once this is done, we find that both the the high-field second-order transition is removed, and that the degeneracy between the two mid-field hexagonal VLs is removed so that only one orientation is expected.

We see that the predictions of the Kogan nonlocal theory depend sensitively on the relationship between the anisotropy of the cores and that of the penetration depth. In YBa$_{2}$Cu$_{3}$O$_{7}$, we may expect the coherence length of the carriers localized in the CuO chains to be quite different in value and degree of anisotropy from the properties of the carriers in the CuO$_{2}$ planes. In addition, Kogan's theory contains no gap anisotropy, and the VL core structure in YBa$_{2}$Cu$_{3}$O$_{7}$ must be affected by the positions of the gap nodes, which will also introduce a nonlocal anisotropy.~\cite{Fra97,Suz10} We can be fairly sure that these effects will remove the degeneracy between the two different orientations of distorted hexagonal VL that are predicted at each field for the twofold symmetric case. Our experimental results show that these two different orientations exist at fields above and below $\sim$2.1~T [Fig~\ref{fig:opening_angles}]; it is possible that a field-dependence of the core anisotropy could lead to the observed field-dependent transition between them. Nevertheless, in order to apply the twofold symmetric form of Kogan's non-local theory to observations real materials, further extensions, perhaps in the form of additional higher-order anisotropy terms, are required. This conclusion is strengthened by the fact that we do not observe the very low field $\sim$45$^{\circ}$ rotated VL (which is apparently expected for all values of anisotropy~\cite{Suz10}). Also the transition to our high-field phase is first-order, again in disagreement with simple nonlocal theory.~\citep{Not04}

\begin{figure}
\includegraphics[width=0.4\textwidth]{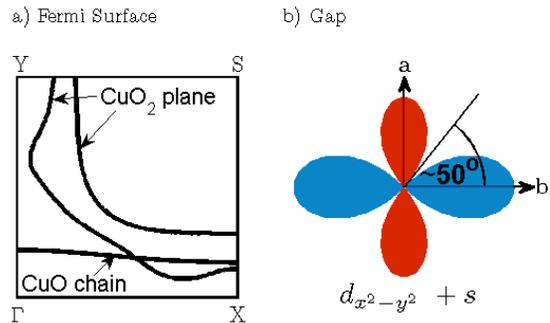}
\caption{(Color online) The zero field in-plane anisotropies of superconducting YBa$_{2}$Cu$_{3}$O$_{7}$ that are expected to influence the VL properties. a) A schematic of the irreducible quadrant of the first Brillouin zone showing the main Fermi surfaces at $k_{z}=0$.~\citep{Pic90,And95} b) The predominantly $d_{x^{2}-y^{2}}$ order parameter combined with the $s$-wave admixture suggested by phase-sensitive measurements.~\citep{Kir06}}
\label{fig:anisotropy}
\end{figure}

Despite the noted shortcomings of the Kogan theory in its current form, we nevertheless emphasize certain details within the predictions that may be compatible with our data. Firstly, we have revealed that both a low field and first-order VL reorientation transition may be expected for the twofold symmetric case. After appropriate modification of the theory, it may be turn out that this transition will involve a 90$^{\circ}$, and not a $\sim$45$^{\circ}$, re-orientation of the primitive cell, and thus be more directly analogous to our observed transition between the LFS and IFS phases. Secondly, we find that the high field VL is expected to evolve \emph{towards} a distorted square-like structure that is analogous to the behavior observed in the HFS phase. While we have clarified that in a \emph{real} twofold symmetric system the predicted high field second-order transition can not occur, the important detail is that at high field the vortex nearest neighbors are expected to \emph{approach} directions parallel to minima in the Fermi velocity.~\cite{Kog97a} This observation is compatible with calculations of the zero field Fermi surface of YBa$_{2}$Cu$_{3}$O$_{7}$~\citep{Pic90,And95} shown schematically in Fig.~\ref{fig:anisotropy}~(a). The dominant bands are large square-like contours associated with the CuO$_{2}$ planes that are distorted so that the Fermi velocity is larger along $k_{x}$ and $k_{y}$ than along the $\{110\}$ directions. This suggests that the structural orientation of the HFS may be determined by the anisotropy of these Fermi surface sheets.~\citep{Kog96,Kog97a,Kog97b}

On the other hand, it may be expected that the strong order parameter anisotropy is more important at high fields than Fermi surface anisotropy. Since the Kogan nonlocal theory does not account for any superconducting gap anisotropy, we now consider how this anisotropy may influence the VL structure. Both the high field transition and the tendency towards a square VL structure may be interpreted as caused by the increasing importance with field of the predominantly $d$-wave superconducting gap. This is a common prediction made using other nonlocal models~\citep{Ber95,Xu96,Aff97,Fra97,Shi99} and more detailed microscopic theories.~\citep{Ich99,Nak02,Suz10} According to the predictions of microscopic theory,~\citep{Ich99,Nak02,Suz10} the $\mathcal{F}$ of the high field square structure is determined by the fourfold vortex core anisotropy arising from the $d$-wave order-parameter. The common prediction of all theories is that the vortex nearest neighbors of the square VL are expected to lie parallel to the nodal direction of the order parameter.~\citep{Aff97,Fra97,Ber95,Xu96,Ich99,Shi99,Nak02,Hia08,Suz10} For YBa$_{2}$Cu$_{3}$O$_{7}$ at zero field, phase-sensitive measurements show that the nodes are not at 45$^{\circ}$ to the crystal axes (as expected for a pure $d_{x^{2}-y^{2}}$ superconductor), but at $\pm50^{\circ}$ to the \textbf{b}-axis~\citep{Kir06} as shown in Fig~\ref{fig:anisotropy}~(b). The smooth variation of the angle $\nu$ in the HFS phase towards the square value of 90$^{\circ}$ could reflect a field-driven change in the nodal positions of the gap function. For the case of YBa$_{2}$Cu$_{3}$O$_{7}$, this might indicate a change of the admixture which causes the deviation of the gap symmetry from pure $d_{x^{2}-y^{2}}$. This hypothesis could be tested via high-field STM measurements.

While we may make qualitative comparisons between our structural data and the available theoretical work, a full understanding of our observations requires inclusion of the orthorhombic system symmetry, and disentangling the expected effects due to the anisotropies of both the band structure and the nodal gap. This is possible in materials where both the band structure and gap anisotropies are expected to yield VL structures of different orientations, e.g. in La$_{1.83}$Sr$_{0.17}$CuO$_{4+\delta}$.~\citep{Gil02,Nak02} In YBa$_{2}$Cu$_{3}$O$_{7}$ however, these two anisotropies are expected theoretically to yield similar VL structures.~\citep{Nak02,Suz10} A more complete understanding of the new structural data presented here calls for extensions to the currently available theoretical work. In particular, our discussion of the Kogan nonlocal theory~\citep{Kog97a,Kog97b} indicates that instead of relying on approximate approaches, a better proposition is to attempt full first-principles numerical calculations.

\subsection{Vortex Lattice Form Factor at 2~K}
\label{sec:3Res_VL_FF_2K}

\subsubsection{Field-dependence of the form factor}

\begin{figure}
\includegraphics[width=0.5\textwidth]{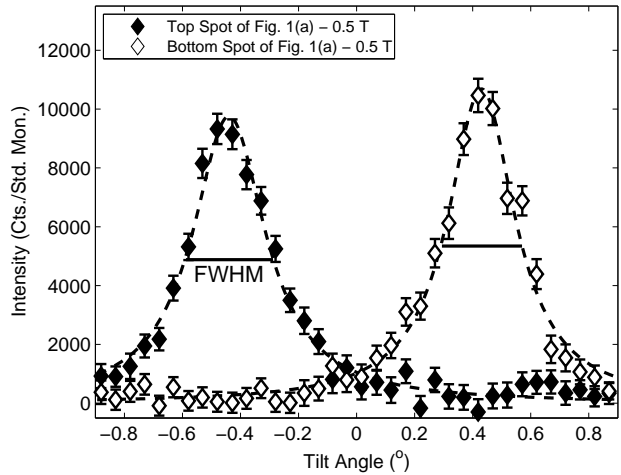}
\caption{The angular dependence of the diffracted intensity (rocking curve) for the top and bottom Bragg spots of the diffraction pattern shown in Fig.~\ref{fig:structures}~(a). The dashed lines are fits of Lorentzian lineshapes to the data, and the solid lines indicate the full width at half maximum (FWHM).}
\label{fig:rocking_curve}
\end{figure}

Within the mixed state, the local field is expressed as a sum over its spatial Fourier components at the various different scattering vectors $q$ belonging to the two-dimensional reciprocal lattice of the VL. The form factor at a wavevector $q$ is the magnitude of the Fourier component $F(q)$, and its value is obtained from the integrated intensity, $I_{q}$, of a VL Bragg reflection as the VL is rotated through the diffraction condition. $I_{q}$ is related to the modulus squared of the form factor, $|F(q)|^{2}$, by~\cite{Chr77}
\begin{equation}\label{1}
    I_{q} = 2\pi V\phi \left(\frac{\gamma}{4}\right)^{2} \frac{\lambda_{n}^{2}}{\Phi_{0}^2 \; q}|F(q)|^{2}.
\end{equation}
Here, $V$ is the illuminated sample volume, $\phi$ is the incident neutron flux density, $\lambda_{n}$ is the neutron wavelength, $\gamma$ is the magnetic moment of the neutron in nuclear magnetons ($=1.91$), and $\Phi_{0} = h/2e$ is the flux quantum. Eq.~\ref{1} shows that the modulus form factor, $|F(q)|$, for a specific Bragg spot is obtained from both $I_{q}$ and the magnitude $q$, with all other variables constant. In order to record $I_{q}$ experimentally, the sample and cryomagnet are rotated together through an angular range that moves a reciprocal lattice vector through the Bragg condition at the detector. Fig.~\ref{fig:rocking_curve} shows representative scans of the angular dependence of the diffracted intensity (rocking curves) for the top and bottom spots with $\textbf{q}\parallel \textbf{a}^{\ast}$ that form part of the diffraction pattern shown in Fig.~\ref{fig:structures}~(a). The quantity $I_{q}$ is obtained by integrating the area underneath the lineshape that is used to fit the rocking curve. When appropriate, $I_{q}$ is corrected by the cosine of the angle between the scan direction and $q$ (the Lorentz factor).~\citep{Squ78}



\begin{figure}
\includegraphics[width=0.5\textwidth]{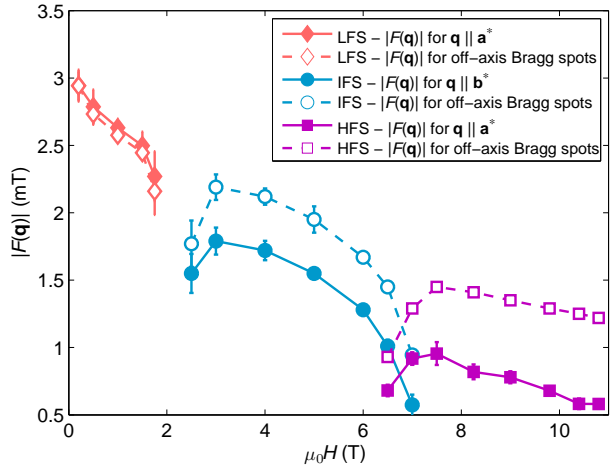}
\caption{(Color online) The field-dependence at 2~K of the observed form factors $F(\textrm{\textbf{q}})$. Across the entire field range, we distinguish between the different types of Bragg spot according to the different magnitudes of the \textbf{q}-vectors for each type. Each datapoint represents the average form factor for a certain spot type.}
\label{fig:form_factor_2K}
\end{figure}

In Fig.~\ref{fig:form_factor_2K}, we show our measurements at 2~K of the VL form factor across the entire field range up to 10.8~T. At each field, we distinguish between form factor values obtained from different types of Bragg spots, and each value represents a mean taken over equivalent spots. Close to the first-order VL structure transitions, the form factor values are noticeably lower than those expected from an extrapolation of `pure phase' data recorded away from these field regions. This is owing to the strong field-dependence of the occupation fraction of each of the two relevant structures close to the transitions, and is an effect not accounted for by Eq.~\ref{1}. For the low field transition, one sees reduced form factor values only at fields of 1.75~T and 2.5~T. For fields within this range, we were unable to measure reliably the integrated intensities for the Bragg spots due to a disordering of the VL. For the high field transition, reduced form factor values are seen at 6.5~T and 7~T. In this case, the two structures co-exist cleanly and it is possible to account for all of the diffracted intensity. However, when modeling our form factor data, we only include data obtained at fields where there is just one VL structure type in the sample.

\begin{figure}
\includegraphics[width=0.5\textwidth]{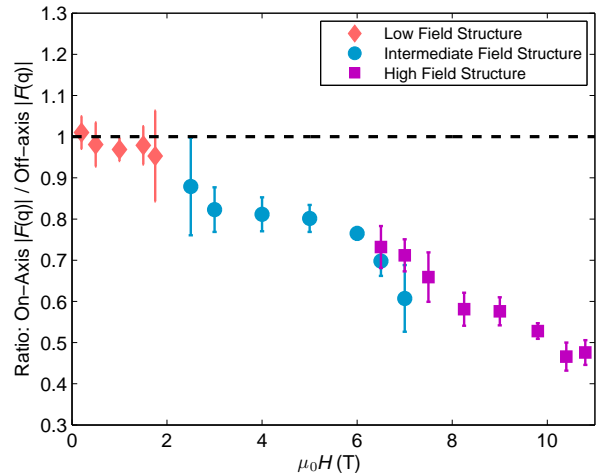}
\caption{(Color online) The field-dependence at 2~K of the ratio between on-axis and off-axis form factors. For the LFS and HFS phases, the on-axis spots lie along $\textbf{a}^{\ast}$, whilst for the IFS phase, the on-axis spots lie along $\textbf{b}^{\ast}$. The black dashed line indicates a ratio of unity, which is expected in the anisotropic London limit.}
\label{fig:form_factor_ratios_2K}
\end{figure}

The data shown in Fig.~\ref{fig:form_factor_2K} indicate that with increasing field, an increasingly significant anisotropy emerges between the form factors of the two types of Bragg spot. This is explicitly shown in Fig.~\ref{fig:form_factor_ratios_2K}, where for each VL structure type we present the field-dependence of the form factor ratio $\left|F(\textbf{q}~ \textrm{on-axis})\right|/\left|F(\textbf{q}~ \textrm{off-axis})\right|$. We see that this defined ratio varies monotonically over the much of the field range, even though the primitive cell of the IFS phase is orthogonal to that of the LFS and HFS phases. For the LFS phase, the data in Fig.~\ref{fig:form_factor_ratios_2K} show that all the spots have the same form factor within the experimental uncertainty, thus giving a form factor ratio close to unity. This is as expected from London theory, even for the case of an anisotropic VL. In moving from the isotropic case to the anisotropic case appropriate for YBa$_{2}$Cu$_{3}$O$_{7}$, $F\left(q\right)$ changes according to

\begin{eqnarray}\label{2}
F\left(q\right)=\frac{\langle B\rangle}{1+q^{2}\lambda^{2}}\rightarrow
\frac{\langle B\rangle}{1+\textrm{q}_{x}^{2}\lambda_{a}^{2}+\textrm{q}_{y}^{2}\lambda_{b}^{2}}.
\end{eqnarray}

Here, $\langle B\rangle$ is the average internal induction (which within experimental error is always $=\mu_{0}H$ due to the plate-like crystals and large $\kappa$) and $\lambda_{i}$ is the penetration depth along the directions $i$.~\citep{Not01} The terms $q_{x}$ and $q_{y}$ represent the $x$ and $y$ components of the \textbf{q}-vector as viewed in the diffraction patterns shown in this paper; thus $q_{x}$ is parallel to $\textbf{b}^{\ast}$, and $q_{y}$ parallel to $\textbf{a}^{\ast}$. In this case, it may be shown~\citep{Thi89} that the \textbf{a}-\textbf{b} anisotropy in the \textbf{q}-vectors in this distorted hexagonal structure exactly cancels the penetration depth anisotropy, giving the observed result. Another prediction of London theory is that the form factor is independent of field for $B\gg B_{\textrm{c1}}$ (where the 1 in the denominators of Eq.~\ref{2} may be neglected). The fall-off shown in Fig.~\ref{fig:form_factor_2K} will be discussed in the following section. For the IFS phase, the observed form factor anisotropy unusually has $\left|F(\textbf{q}\parallel \textbf{b}^{\ast})\right|<\left|F(\textbf{q}~ \textrm{off-axis})\right|$ giving a \emph{smaller} value of $\left|F(q)\right|$ for the Bragg spot with the \emph{shorter} \textbf{q}-vector.

\begin{table*}[t]
\caption{\label{tab:table1}A summary of the returned parameters obtained after fitting the 2~K form factor data with both the London and Clem models (see text for details of the fitting). The values of $\chi^{2}$ are those obtained after each step of the analysis procedure.}
\begin{ruledtabular}
\begin{tabular}{ccccccc}
 Model & $\lambda_{a}$~(nm)& $\lambda_{b}$~(nm) & $\xi_{a}$~(nm) & $\xi_{b}$~(nm) & $\chi^{2}$~(Step~1) & $\chi^{2}$~(Step~2) \\ \hline
 London & 138(2) & 107(1) & 3.04(4) & 3.54(11) & 1.46        & 1.92       \\
 Clem   & 132(5) & 103(1) & 2.72(13)  & 2.49(36)  & 4.40        & 13.27 \\
\end{tabular}
\end{ruledtabular}
\end{table*}


\subsubsection{Modeling the field-dependence of the VL form factor}
The simplest analytic models that are most commonly used to interpret form factor data are those based on applying corrections to local electrodynamics suggested by Ginzburg-Landau (GL) theory.~\cite{Cle75,Hao91,Yao97} Although GL theory is only strictly valid close to $T_{c}$, the results of more exact numerical calculations~\cite{Bra97,Bra03,Ich99} show that such simple models~\cite{Cle75,Hao91,Yao97} provide a reasonable description of the mixed state at lower $T$. In the isotropic case, the model is
\begin{eqnarray}\label{3}
F\left(q\right)=\frac{\langle B\rangle \textrm{exp}\left(-c q^{2}\xi^{2}\right)}{1+q^{2}\lambda^{2}}.
\end{eqnarray}
which can be considered as a refined London model with a gaussian cut-off term that accounts for the finite size of vortex cores. Here, $\xi$ is the in-plane GL coherence length and $\lambda$ is the in-plane penetration depth. The constant $c$ is a parameter expected to have a value between 1/4 and 2.~\citep{Yao97} A quantitative comparison between the predictions of the above London model and those of numerical calculations carried out within the Eilenberger formalism~\citep{Ich99,Ich07} suggest that over our field range, and at low $T$, an appropriate value for $c$ is 0.44.~\citep{Bow08} We note that this value lies close to that of $\frac{1}{2}$, which has been used in successful form factor analyses for other superconductors.~\citep{Cub07,Den09}

Other authors have used the Clem model,~\citep{Cle75} which is obtained from a variational solution to the GL equations. The model is:
\begin{eqnarray}\label{4}
F\left(q\right)=\langle B\rangle\frac{gK_{1}\left(g\right)}{1+q^{2}\lambda^{2}},\quad
g=\sqrt{2}\xi_{GL}\left(q^{2}+\lambda^{-2}\right)^{1/2}
\end{eqnarray}
where $K_{1}$ is a modified Bessel function of the second kind. This model appears advantageous in that there is no ill-defined $c$ factor within the core-correction term $gK_{1}(g)$ as there is in Eq.~\ref{3}, but it relies on the GL equations which are only valid close to $T_{\textrm{c}}$. As we will show, this approach gives a much poorer fit to our data than the refined London model described by Eq.~\ref{3}.

Both of these models can be extended in the same way in order to account for the biaxial anisotropy of YBa$_{2}$Cu$_{3}$O$_{7}$. For example, the refined London model (Eq.~\ref{3}) can be re-written in the convenient form,
\begin{eqnarray}\label{5}
F\left(\textrm{q}\right)=\frac{\langle B\rangle \textrm{exp}\left(-0.44\left(\textrm{q}_{x}^{2}\xi_{b}^{2}+\textrm{q}_{y}^{2}\xi_{a}^{2}\right)\right)}{\textrm{q}_{x}^{2}\lambda_{a}^{2}+\textrm{q}_{y}^{2}\lambda_{b}^{2}}.
\end{eqnarray}
Here $\xi_{i}$ represent the GL coherence lengths along the directions $i$. According to the extended models, in general the form factor is expected to depend on the four parameters $\lambda_{i}$ and $\xi_{i}$. However, the number of free parameters can be reduced by fitting the model independently for certain types of spots. For spots with $\textbf{q}\parallel \textbf{a}^{\ast}$, terms dependent on $q_{x}$ vanish, as do those dependent on $q_{y}$ for spots with $\textbf{q}\parallel \textbf{b}^{\ast}$. Taking advantage of this, we extract values for $\lambda_{i}$ and $\xi_{i}$ from our form factor data using the following analysis procedure:
\begin{list}{Step}{}
\item 1. We fit the form factor data for Bragg spots with $\textbf{q}\parallel \textbf{a}^{\ast}$ in order to obtain $\lambda_{b}$ and $\xi_{a}$.
\item 2. The values of $\lambda_{b}$ and $\xi_{a}$ found in Step~1 are held constant, and we fit the data for the spots with an off-axis \textbf{q}-vector in order to obtain $\xi_{a}\left(0\right)$ and $\lambda_{b}\left(0\right)$.
\end{list}
For both steps, the values of $q_{x}$ and $q_{y}$ at each field, and their associated errors, are those determined experimentally. We also see that the fit in each step is sensitive to just two fitting parameters.


\begin{figure}
\includegraphics[width=0.5\textwidth]{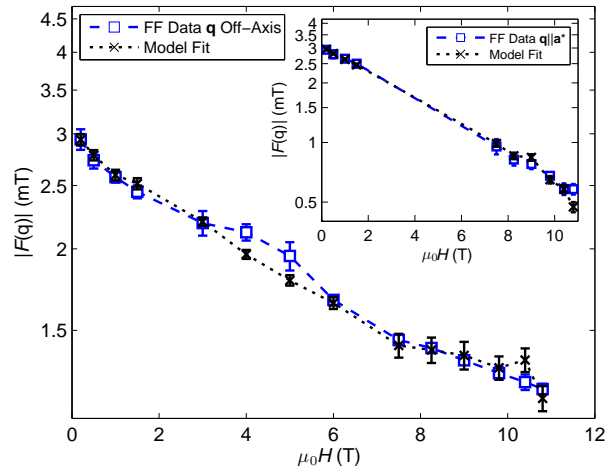}
\caption{(Color online) A semi-logarithmic graph showing the field-dependence of the VL form factor for Bragg spots with a \textbf{q}-vector lying off-axis. The fit of the data to the London model is also shown. The inset shows the model fit of the form factor data with $\textbf{q}\parallel \textbf{a}^{\ast}$.}
\label{fig:London analysis}
\end{figure}


In Fig.~\ref{fig:London analysis} we show the fits of the form factor data to the London model. The first step of the analysis procedure is shown in the inset, and yields parameters of $\lambda_{b}=107(1)$~nm and $\xi_{a}=3.04(4)$~nm. The reasonable fit quality is indicated by $\chi^{2}=1.46$. In the main panel we show the results of the second fit, with the fitted parameters $\lambda_{a}=138(2)$~nm and $\xi_{b}=3.54(11)$~nm, and $\chi^{2}=1.96$. As can be seen in Table~\ref{tab:table1}, using the same analysis procedure with the Clem model provides a poorer description of the data than the London model. Attempts were made to fit the data using a version of the Clem model that includes minor corrections considered appropriate for high-$T_{c}$ superconductors.~\citep{Hao91} However, these did not generate an improvement of the fit.


We now discuss the results obtained on fitting the form factor data to the London model (Eqn~\ref{5}). The values returned for $\lambda_{a}$ and $\lambda_{b}$ are in reasonable agreement with those determined by previous SANS studies on optimally- and over-doped YBa$_{2}$Cu$_{3}$O$_{7-\delta}$,~\citep{Joh99} though overall our values are a little shorter. We find better agreement between our values and those found in a recent $\mu$SR study on a clean sample of close to optimally-doped YBa$_{2}$Cu$_{3}$O$_{6.92}$.~\citep{Kie10} This indicates the state of our sample to be of higher quality and more completely oxygenated compared to the sample studied previously by SANS.~\citep{Joh99} On the other hand, the values returned for $\xi_{a}$ and $\xi_{b}$ are likely to be overestimates. Using these values we can estimate the upper critical field to be of order $\sim30$~T, which is clearly too low for YBa$_{2}$Cu$_{3}$O$_{7}$. One possibility for the over-estimation of $\xi_{i}$ is that VL disorder may reduce the measured intensity in the form of a low $T$ `static' Debye-Waller (DW) factor. Such a factor would enter the numerator of Eq.~\ref{5} as $\textrm{exp}\left(-q^{2}\langle u^{2}\rangle /4\right)$, where $\langle u^{2}\rangle$ is the root-mean-square displacement of a vortex along the direction of $q$. By inspection it is clear that this term has precisely the same effect on the VL form factor as the Gaussian cut-off term that accounts for the finite vortex core size, and that both of these terms become increasingly important at larger $q$. As a consequence of this, we are unable to disentangle the effects of $\xi$ or $\langle u^{2}\rangle$. However, bearing in mind that in high-$T_{c}$ materials both the value of $\xi$ is short, and the vortices are `soft' (small values of the shear modulus and the tilt modulus at small distances~\citep{Bra95}) it is unsurprising that disorder in the form of a DW factor may become important. This is particularly the case at higher fields where the VL spacing is just a few times the value of $\xi_{i}$.

An alternative possibility is that any of $\xi_{i}$ or $\lambda_{i}$ exhibit a significant field-dependence over our field range. Previous high field $\mu$SR studies on YBa$_{2}$Cu$_{3}$O$_{6.95}$ report both a strong increase with field of the extracted in-plane penetration depth~\citep{Son97,Son99} attributed to non-local effects,~\citep{Ami98,Ami00} and also a strong decrease with field of the vortex core size $\xi$ attributed to a field-induced quenching chain superconductivity.~\citep{Atk08} It should be noted that in the $\mu$SR analysis~\citep{Son99} an undistorted hexagonal VL was assumed, which we have shown is not the case. Also, the Clem model was used in a $T$ range where its validity is questionable. Within the framework of our form factor model, unconventional field-dependences of $\lambda_{i}$ and $\xi_{i}$ may offset one another to give alternative fits to the data. However, we hesitate to fit for field-dependent length-scales as this amounts to inserting unknown free parameters.

\begin{figure*}
\includegraphics[width=0.45\textwidth]{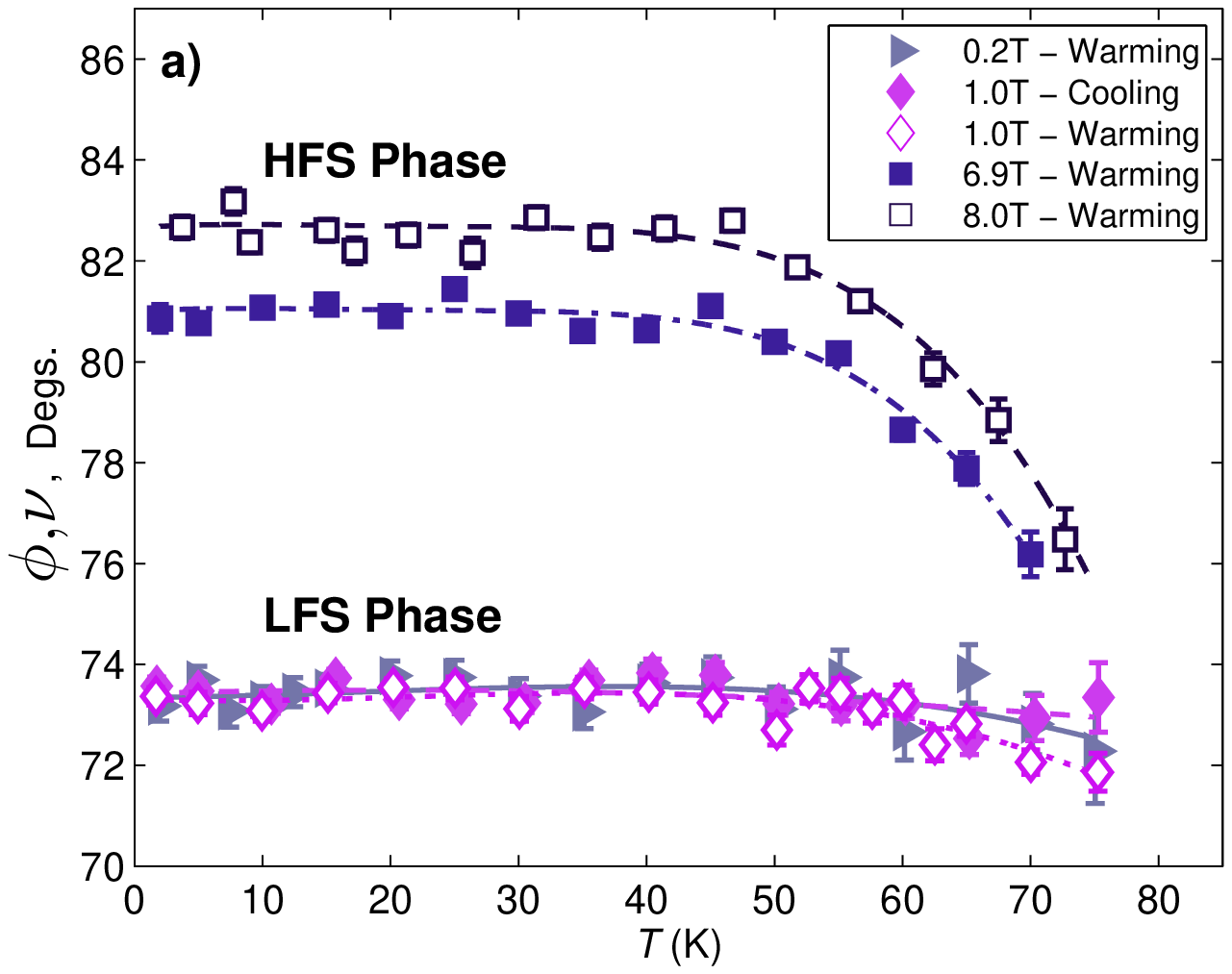}
\includegraphics[width=0.45\textwidth]{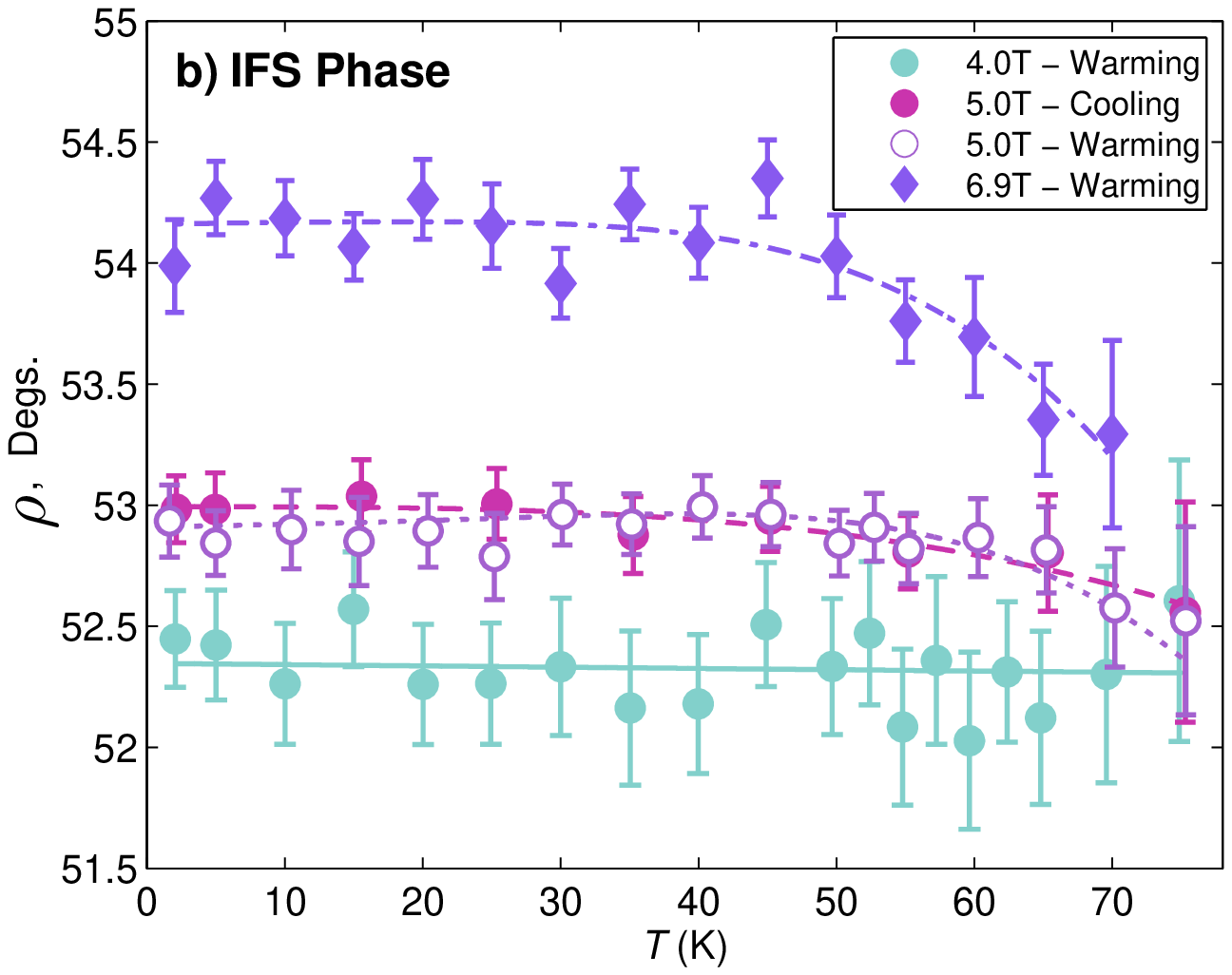}
\caption{(Color online)
The temperature-dependence of the VL structure opening angles (a) $\phi$ and $\nu$ for the LFS and HFS phases respectively, and (b) $\rho$ for the IFS phase. The indicated angles correspond to those in Fig.~\ref{fig:structures}. In both panels, the lines passing through for the data at each field are guides to the eye. Data are shown only for temperatures where the spot position on the detector could be determined accurately. Error bars not visible can be considered of order the size of the datapoint.}
\label{fig:T-dep_Structure}
\end{figure*}


In spite of our conclusion that the values for $\xi_{i}$ are too large, the returned values for $\lambda_{a}=138(2)$~nm and $\lambda_{b}=107(1)$~nm are likely to represent the intrinsic bulk values. Within the local London model described by Eq.~\ref{5}, the values of $\lambda_{i}$ are determined by the form factor magnitude at lower fields, and are hence rather insensitive to any exponential terms in the numerator. From our analysis we find that $\gamma_{\lambda}=\lambda_{a}/\lambda_{b}=1.29(2)$, a value which is in good agreement with that obtained directly from the measure of the distortion of the hexagonal structure shown in Fig.~\ref{fig:hex_distortion}. This strongly indicates that local theory is at least able to describe the anisotropy of the low field VL. In addition, the evolution of the form factor anisotropy with field [Fig.~\ref{fig:form_factor_ratios_2K}] could arise, at least in part, if the value of $\gamma_{\lambda}$ \emph{does not} fall significantly with increasing field. In this case, the form factor anisotropy emerges as a consequence of the changes in VL structure and the associated changes in \textbf{q}, rather than due to anisotropic field-induced variations in the fundamental length-scales.

We emphasize that conclusions drawn from form factor analyses such as ours, and also those of $\mu$SR data, need to be considered with great care, and provide only an approximate picture of the mixed state. A more thorough understanding of our form factor data might be obtained by harnessing the power of a microscopic theory,~\cite{Eil68,Ich99,Ich07} where additional effects such as those due to non-locality and order-parameter anisotropy are included by default.

\subsection{Measurements at Higher Temperatures}
\label{sec:3Higher_Ts}
\subsubsection{VL Structure at Higher Temperature}

Our $T$-dependent measurements were carried out by warming from, or cooling into, an oscillation field-cooled state. While the $T$ was changing, the field was oscillated using the OFC procedure. Upon reaching the desired $T$, the field was held stationary at the intended value before starting the measurement. In Fig.~\ref{fig:T-dep_Structure} we present $T$-dependent measurements of the VL structure opening angles defined in Fig.~\ref{fig:structures} at various magnetic fields. We checked for hysteretic behavior by performing both warming and cooling $T$ scans at 1.0~T [Fig.~\ref{fig:T-dep_Structure}~(a)] and 5.0~T [Fig.~\ref{fig:T-dep_Structure}~(b)]. At these two fields, there is no discernible difference in the precise VL structure between warming and cooling, thus showing that hysteresis effects on the VL structure can be neglected in our sample.

Fig.~\ref{fig:T-dep_Structure}~(a) reveals that for fields within the LFS phase, the VL structure remains essentially $T$-independent. Only when $T\rightarrow T_{c}$ is there any measurable variation in the opening angle $\phi$, though this variation remains small. Data obtained in the IFS phase [Fig.~\ref{fig:T-dep_Structure}~(b)] indicate that this weak $T$-dependence of the VL structure persists until fields of up to 5.0~T. However,more pronounced $T$-dependences emerge at higher fields. At 6.9~T, a field where the IFS and HFS phases co-exist at 2~K, there is clear $T$-induced reduction in the values of both $\rho$ and $\nu$ as $T\rightarrow T_{c}$. At yet higher fields in the HFS phase, the data at 8.0~T show the high $T$ reduction of $\nu$ to become increasingly pronounced. Remarkably, the data in Fig.~\ref{fig:T-dep_Structure} reveal that for no field does adjusting the $T$ cause a transition between \emph{different} VL structures. Even at 6.9~T, the data show that the low $T$ co-existence of the IFS and HFS phases persists over the entire $T$ range. All of these observations suggest that the phase boundaries between different structure types are unusually $T$-independent.

It also remains to be explained why any variation in the VL structure occurs only at high $T$, and at relatively high fields. The former observation suggests that, in spite of utilizing the OFC procedure, the VL structure is still `frozen-in' and pinned at a relatively high $T_{\textrm{irr}}$ of $\simeq$50~K. If so, any variation of the VL structure with $T$ would only occur in the reversible region, when the vortices are depinned and able to adopt the equilibrium structure. For fields below $\simeq$~5.0~T, there is little variation of the structure at higher $T$, suggesting that there will not be a significant difference between the precise VL structure frozen-in at $T_{\textrm{irr}}$, and that expected for low $T$. However, at higher fields a larger difference emerges between the VL structure frozen-in, and that which may be expected at lower $T$ by extrapolating the high $T$ behavior.

The high $T$ variation of the VL structure can be understood within the framework of non-local theory,~\citep{Kog97a,Fra97,Suz10} where the non-locality can arise either from Fermi surface anisotropy or gap structure. A common prediction is for non-local effects to become suppressed as $T\rightarrow T_{\textrm{c}}$. This is most clear to see at fields within the HFS phase, where on the approach to $T_{c}$ we observe the values of $\nu$ to approach those values of $\phi$ observed in the LFS phase where we believe that anisotropy due to \emph{local} effects is dominant. Non-local effects are also suppressed by $T$ within the IFS phase, and the high-$T$ variation in $\rho$ shows that the VL tends to become slightly \emph{more} anisotropic (larger value of $\eta$) close to $T_{\textrm{c}}$. In this case, the suppression of non-locality allows the VL to adopt a hexagonal coordination that is more distorted. A common interpretation of these data is that the high $T$ or low field anisotropy factor ($\sim$1.29) just arises due to the band mass anisotropy, and can be understood with local anisotropic London theory.

We now consider the remarkable $T$-independence of the VL structure transition fields. It is expected that non-local effects fall off with increasing $T$ but not to zero.~\citep{Kog97a,Fra97} Clearly at $T$=$T_{\textrm{c}}(H)$, these effects remain strong enough to control the orientation of the VL primitive cell, and it appears that the crossover fields remain essentially $T$-independent below this. However, in the IFS and HFS phases the \emph{shape} of the VL evolves away from square structures with increasing $T$. This tendency is also consistent with that expected if the VL structures are stabilized by the $d$-wave gap anisotropy.~\citep{Ich99,Hia08} With increasing $T$, the influence of the order parameter anisotropy on the VL structure is expected to be suppressed, as quasiparticles that previously occupied states near the nodal positions adopt a greater angular spread in reciprocal space. However, if the $d$-wave gap anisotropy is responsible for \emph{driving} the VL structure transitions, and controlling the primitive cell \emph{orientation}, the influence of this anisotropy must persist to surprisingly high $T$.

\subsubsection{VL Form Factor at Higher Temperatures}
It seems that the precise VL structure is frozen-in on cooling below $T_{\textrm{irr}}$; this provides the opportunity to record the $T$-dependence of the form factor at low fields without requiring an excessive amount of neutron beamtime. Typically, carrying out these measurements is a lengthy process; at each $T$, full rocking curves such as those shown in Fig.~\ref{fig:rocking_curve} should be obtained for each Bragg spot. However, in our case where the precise VL structure is weakly $T$ dependent, the same information can be obtained by counting solely at the peak of the rocking curve. Knowing this peak intensity, along with the FWHM of the rocking curve and the \textbf{q}-vector allows a calculation of the form factor. This approach assumes that the FWHM of the rocking curve remains $T$ independent. This assumption was checked within each $T$ scan. While at most temperatures, measurements were carried out solely at the Bragg angle, at certain temperatures we carried out full rocking curve measurements. These measurements provide both a check of the assumed $T$ independence of the rocking-curve FWHM, and also full measures of the form factor to be compared with those obtained from the fixed angle measurements.

\begin{figure*}
\includegraphics[width=0.45\textwidth]{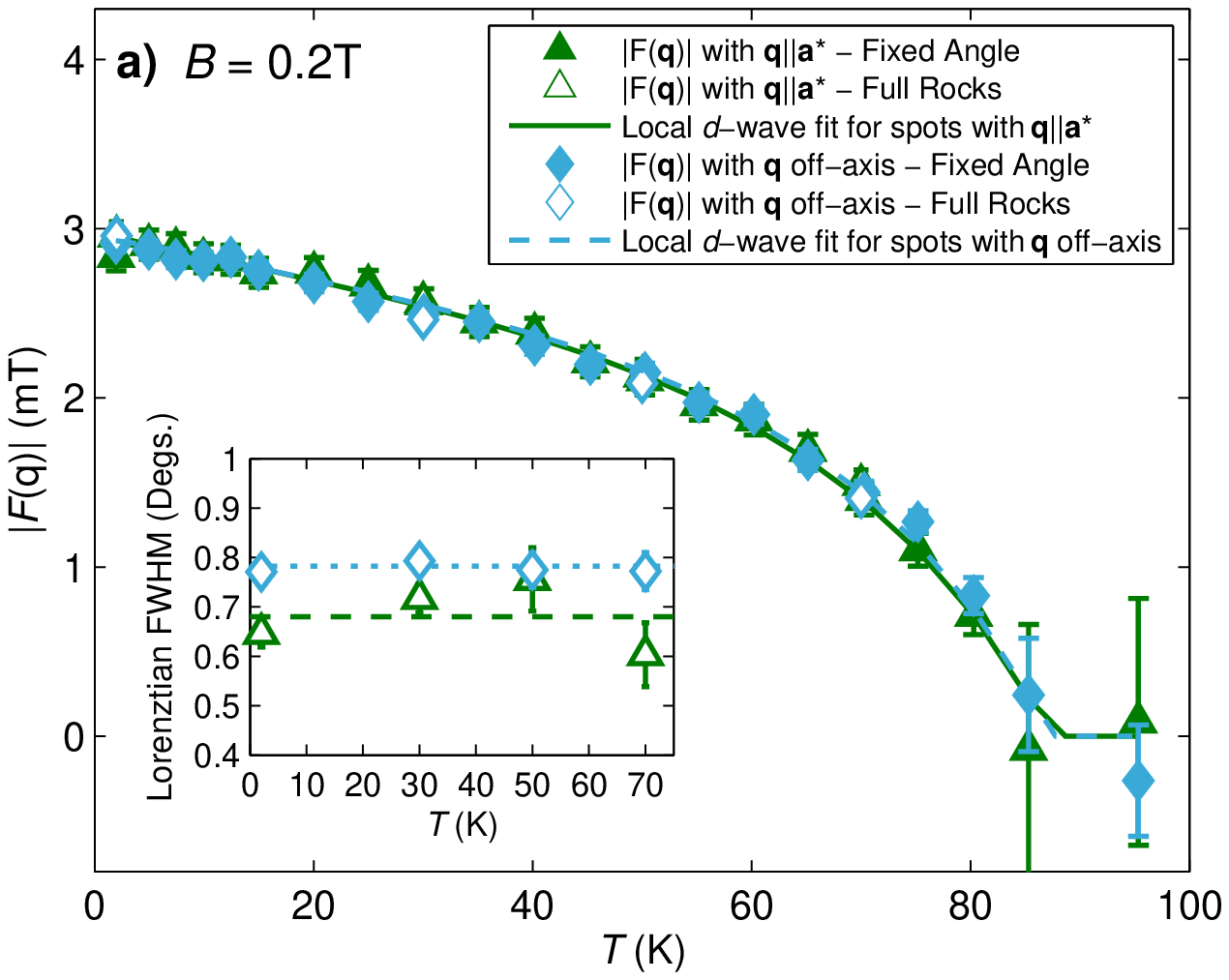}
\includegraphics[width=0.45\textwidth]{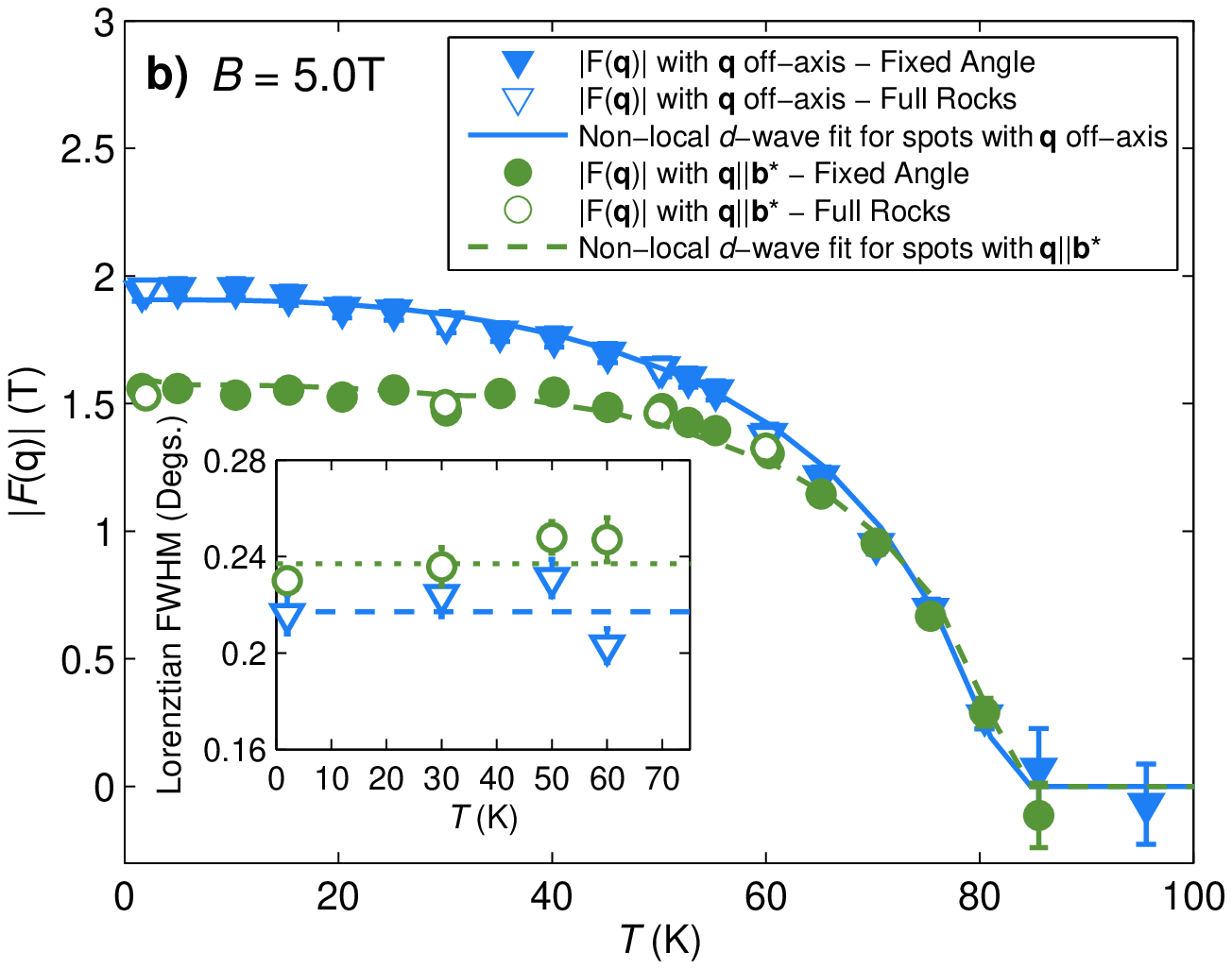}
\caption{(Color online)
The temperature dependence of the form factor $|F(q)|$ for the two types of Bragg spot in (a) the LFS at 0.2~T, and (b) the IFS at 5.0~T. For each scan, we performed an OFC procedure down to base temperature, and then measured on heating the sample. In (a), the solid and dashed lines show fits to the data made using both a minimal and local $d$-wave model. In (b), solid and dashed lines show fits to the data after extending the local $d$-wave model used in (a) with a non-local correction. Both models are described in the text. The inset of each figure shows the temperature-dependence of the rocking curve FWHM as obtained from the rocking curve measurements. The dashed and dotted lines, respectively, indicate the mean value of the FWHM used to normalize the form factor values of the fixed angle data (see text for details).}
\label{fig:T-deps}
\end{figure*}


In Fig.~\ref{fig:T-deps} we present the $T$ dependence of the form factor in applied fields of 0.2~T (Fig.~\ref{fig:T-deps}~(a)) and 5.0~T (Fig.~\ref{fig:T-deps}~(b)). At each field, and for each type of Bragg spot, we show both the values obtained by measuring at fixed rotation angle and those obtained with the usual full rocking curves. The full rocking curve measurements allow us to plot the $T$ dependence of the FWHM of the rocking curve for each type of Bragg spot. Both the insets to Figs.~\ref{fig:T-deps}~(a) and (b) show that there is no clear systematic $T$ variation. Therefore, we use a mean value for the FWHM of each type of Bragg spot across the entire $T$ range. As seen in Figs.~\ref{fig:T-deps}~(a) and (b), the agreement between the two techniques of obtaining the form factor is good, thus providing confidence in our measurements obtained at fixed rotation angle. We also note that any $T$-induced variation in the precise VL structure is small, and does not extend beyond the in-plane resolution of the instrument. Therefore, at these fields, any change in the VL structure will have a negligible influence on the values of intensity obtained at fixed angle.

In Fig.~\ref{fig:T-deps}~(a) we see that at 0.2~T the form factors for the two types of Bragg spot exhibit a similar $T$-dependence. Interpreting these data requires a knowledge of the $T$-dependences for both $\xi_{i}\left(T\right)$ and $\lambda_{i}\left(T\right)$, each of which require a detailed knowledge of the band structure. While ARPES studies~\cite{Sch98,Zab07} and various calculations have established the main band structure reasonably well,~\citep{Pic90,And95} computing the $T$ dependence of the superfluid density in materials like YBa$_{2}$Cu$_{3}$O$_{7}$ is not trivial since this requires a precise knowledge of the momentum-resolved quasiparticle spectrum. As this information is not readily available, we instead make use of a `minimal model' for the Fermi surface sheets arising from the CuO$_{2}$ planes, and represent them by a single two dimensional quasi-cylindrical sheet, which has some \textbf{a}-\textbf{b} anisotropy.~\citep{Pic90,And95,Sch98,Zab07} Whilst such an approach neglects a possible contribution to the superfluid density due to chain superconductivity, it is expected that the dominant contribution will arise due to the large hole-like sheets centered at the Brillouin zone corner.

According to Eq.~\ref{5}, the form factor $F\left(q, T\right)$ for a Bragg spot will be temperature-dependent according to:

\begin{equation}\label{6}
F\left(\textrm{q}, T\right)=\frac{\langle B\rangle \textrm{exp}\left(-0.44\left(\textrm{q}_{x}^{2}\xi_{b}^{2}\left(T\right)+\textrm{q}_{y}^{2}\xi_{a}^{2}\left(T\right)\right)\right)}{\textrm{q}_{x}^{2}\lambda_{a}^{2}\left(T\right)+\textrm{q}_{y}^{2}\lambda_{b}^{2}\left(T\right)}.
\end{equation}
In the local approximation, the $T$-dependence of the penetration depth can be calculated using the following equation suitable for a two-dimensional and cylindrical Fermi surface,~\citep{Pro06}
\begin{eqnarray}\label{7}
\frac{1}{\lambda_{a/b}\left(T\right)^{2}}=\rho_{s,a/b}\left(T\right)=\qquad\qquad\qquad\qquad\qquad\qquad\qquad \nonumber\\
1-\frac{1}{4\pi T}\int_{0}^{2\pi}\int_{0}^{\infty}\textrm{cosh}^{-2}\left(\frac{\sqrt{\epsilon^{2}+\Delta_{k}^{2}\left(T, \phi\right)}}{2k_{B}T}\right)\textrm{d}\phi\textrm{d}\epsilon.\qquad
\end{eqnarray}
Here, $\phi$ is the azimuthal angle around the cylindrical Fermi surface and $\sqrt{\epsilon^{2}+\Delta_{k}^{2}\left(T,\phi\right)}$ defines the excitation energy spectrum. The gap function $\Delta_{k}\left(T,\phi\right)$ is assumed separable into momentum- and $T$-dependent factors such that $\Delta_{k}\left(T,\phi\right)=\Delta_{k}\left(\phi\right)\Delta_{0}\left(T\right)$ where $\Delta_{k}\left(\phi\right)$ describes the angular-dependent part of the gap function, and $\Delta_{0}\left(T\right)$ the $T$-dependent part. The angular dependence for the gap function can be expressed as $\Delta_{k}\left(\phi\right)=1$ for an $s$-wave gap, or $\Delta_{k}\left(\phi\right)=\textrm{cos}(2\phi)$ for a $d$-wave gap. The $T$-dependence of the gap is calculated using the following expression,~\citep{Gro86} $\Delta_{0}\left(T\right)=\Delta_{0}\left(0\right)\textrm{tanh}\left(1.78\sqrt{T_c/T-1}\right)$, where $\Delta_{0}\left(0\right)$ is the maximum magnitude of the gap at zero $T$. This gives a good approximation to the BCS weak-coupling $T$-dependence, though $\Delta_{0}$ does not need to have the BCS value. We also use this expression for the gap to calculate the $T$-dependence of $\xi_{a,b}\left(T\right)$,
\begin{equation}
\xi_{a/b}\left(T\right)=\xi_{a/b}\left(0\right)\left[\textrm{tanh}\sqrt{1.78\left(T_c/T-1\right)}\right]^{-1}.
\end{equation}
By using the values of $\xi_{i}$ and $\lambda_{i}$ obtained from the 2~K form factor analysis as the zero temperature values, the only fitting parameter for a single gap model is the zero temperature gap magnitude $\Delta_{0}\left(0\right)$.

By making use of a single $d$-wave gap function, Fig.~\ref{fig:T-deps}~(a) shows that the minimal model is able to provide a good description of the data at 0.2~T. The returned values for $\Delta_{0}\left(0\right)$ are $25(2)$~meV for the Bragg spot with $\textbf{q}\parallel \textbf{a}^{\ast}$, and $26(2)$~meV for the off-axis Bragg spots. These values are in reasonable agreement with those found in $\mu$SR experiments on twin-free YBa$_2$Cu$_{3}$O$_{7-\delta}$.~\citep{Kha07}

On the other hand, and again drawing comparison between our results and those in the aforementioned $\mu$SR study,~\citep{Kha07} we do not observe a signature of an inflection point in $\left|F\left(q, T\right)\right|$ at low $T$. This inflection point, which is observed clearly in directly comparable low field measurements of the superfluid density is attributed to a contribution to the superfluid density of a small $s$-wave gap that is quenched with increasing field.~\citep{Kha07} An absence of an inflection point in microwave studies has previously been interpreted as possibly attributable to chain disorder,~\citep{Atk08} which can not be the case in our fully oxygenated sample of YBa$_{2}$Cu$_{3}$O$_{7}$. We also stress that even though the VL is pinned at low $T$, an increase of the superfluid density would be clearly visible in our measurements of the form factor at 0.2~T.

In contrast to the data obtained at 0.2~T, in Fig.~\ref{fig:T-deps}~(b) we see very different behavior at 5.0~T. For each spot type, the low $T$-dependence of the form factor is much weaker, and the minimal model is less able to provide a description of the data. To obtain a better agreement we would need to invoke either a vortex core contraction with \emph{increasing} $T$, or a $T$-dependence to the superfluid density that is more reminiscent of fully-gapped behavior. Both of these scenarios are considered as unlikely.

An alternative route to resolving this issue is to realize that a weak low-$T$ dependence of the form factor is in better agreement with that expected due to a strong nonlocal response of a $d$-wave superconductor under an applied field.~\citep{Kos97,Ami98,Ami00} Amin~\emph{et al.}~\citep{Ami98,Ami00} have studied the effect of a field-induced nonlocal response on the effective penetration depth in YBa$_{2}$Cu$_{3}$O$_{6.95}$, as observed by $\mu$SR.~\citep{Son99} With increasing field, they find that the usual linear low-$T$ dependence of the superfluid density crosses over to a $T^{3}$ dependence below a temperature $T^{\ast}=\Delta_{0}(\xi_{0}/d)\propto\sqrt{H}$, where $d$ is the VL spacing. These results suggest that the weak low $T$-dependence of $\left|F\left(q, T\right)\right|$ we observe may also be attributable to strong nonlocal effects.

Consistent with the theory of Amin~\emph{et al.}~\citep{Ami98,Ami00}, a simple expression that captures the expected behavior of the superfluid density due to $d$-wave non-local effects is
\begin{eqnarray}\label{8}
\frac{1}{\lambda_{a/b}^{nl}\left(T\right)^{2}}=n_{s,a/b}\left(T\right)\qquad\qquad\qquad\qquad\qquad\qquad\qquad\nonumber\\
=1-\left(1-\rho_{s,a/b}\left(T\right)\right)\left(\frac{T_{c}+T^{\ast}}{T_{c}}\right)\left(\frac{T^{2}}{T^{2}+(T^{\ast})^{2}}\right)\qquad
\end{eqnarray}
where $\rho_{s,a/b}$ is the superfluid density as calculated in the local limit (Eq.~\ref{7}), and $n_{s,a/b}$ is the superfluid density when accounting for the non-local effect. By combining Eqs.~\ref{6} and~\ref{8}, it is possible to calculate the temperature dependence of the form factor with a non-local correction to the penetration depth.

A fit to the data was most easily achieved by using pre-determined values for $\Delta_{0}\left(0\right)$ and $T^{\ast}$, and leaving all of $\xi_{i}$ and $\lambda_{i}$ as free parameters. $\Delta_{0}\left(0\right)$ was fixed to be 25.5~meV, the mean value found for the 0.2~T data. The value of $T^{\ast}$ was held at the temperature where linear extrapolations of the low and high temperature portions of the dataset intersect. At 5.0~T, and for both spot types, this intersection temperature is 51.4(5.0)~K. Further analysis showed that the quality of the fit is actually quite insensitive to $T^{\ast}$, and as such its precise value does not significantly affect the outcome.

Similar to the 2~K form factor analysis, the data were fitted using a two step analysis sequence:
\begin{list}{Step}{}
\item 1. We fitted the data for the Bragg spot with $\textbf{q}\parallel \textbf{b}^{\ast}$ in order to obtain values for $\xi_{b}\left(0\right)$ and $\lambda_{a}\left(0\right)$.
\item 2. The values of $\xi_{b}\left(0\right)$ and $\lambda_{a}\left(0\right)$ found in Step~1 are held constant, and we then fit the off-axis data in order to obtain the values of $\xi_{a}\left(0\right)$ and $\lambda_{b}\left(0\right)$.
\end{list}
Fig.~\ref{fig:T-deps}~(b) shows the good fits of the nonlocal model to the data at 5.0~T with the values obtained from the fitting procedure being $\lambda_{a}\left(0\right)$~=~170(4)~nm, $\lambda_{b}\left(0\right)$~=~121(3)~nm, $\xi_{a}\left(0\right)$~=~1.82(18)~nm and $\xi_{b}\left(0\right)$~=~1.55(15)~nm. However, a good fit is only obtained using the nonlocal model when all of $\xi_{i}\left(0\right)$ and $\lambda_{i}\left(0\right)$ are left as free parameters. Other approaches were tried: using the \emph{local} model (Eq.~\ref{6}) with field-dependent length-scales $\xi_{i}\left(0\right)$ and $\lambda_{i}\left(0\right)$; or using the \emph{non-local} model, but keeping the values of  $\xi_{i}\left(0\right)$ and $\lambda_{i}\left(0\right)$ fixed to those obtained from the 2~K form factor analysis. Both of these gave significantly poorer fits.


By comparing the values of $\xi_{i}(0)$ and $\lambda_{i}(0)$ obtained from fit shown in Fig.~\ref{fig:T-deps}~(b), to the values obtained from the field-dependent form factor analysis at 2~K, we might conclude that $\lambda_{i}(0)$ increases with field, while $\xi_{i}(0)$ decreases with field. The signs of these field-induced variations are consistent with those reported by $\mu$SR studies on YBa$_{2}$Cu$_{3}$O$_{6.95}$.~\citep{Son97,Son99} The possible physical reasons for these field-dependences were introduced previously in our discussion of the field-dependence of the form factor at 2~K. As was also mentioned in that discussion, before making firm statements on any field-dependence for either of $\xi_{i}(0)$ or $\lambda_{i}(0)$, the possible role that VL disorder may play on our measurements should be considered. Since our measurements of the rocking curve FWHM shown in the inset of Fig.~\ref{fig:T-deps}~(b) do not exhibit a large $T$-dependence, we may conclude that a term in the form of the static DW factor does not dominate the low $T$-dependence of the VL form factor for wavevectors associated with fields of up to 5.0~T. Nevertheless, more detailed measurements may reveal a slight $T$-dependence of the rocking curve FWHM, and thus provide alternative descriptions of the data than those solely based on field-dependent length-scales. At 0.2~T, ignoring disorder is more easily justifiable, because the DW factor is closer to unity at the shorter wavevectors associated with low fields. In principle, the role of VL disorder can be determined from direct measurements of higher-order spots of the VL because these spots are more sensitive to disorder effects. Unfortunately, due to the long value of $\lambda$ in YBa$_{2}$Cu$_{3}$O$_{7}$, investigating higher-order spots is experimentally challenging. Indeed, within the IFS phase, we were unable to observe any signal due to higher-order spots of the hexagonal VL.

\begin{figure}
\includegraphics[width=0.5\textwidth]{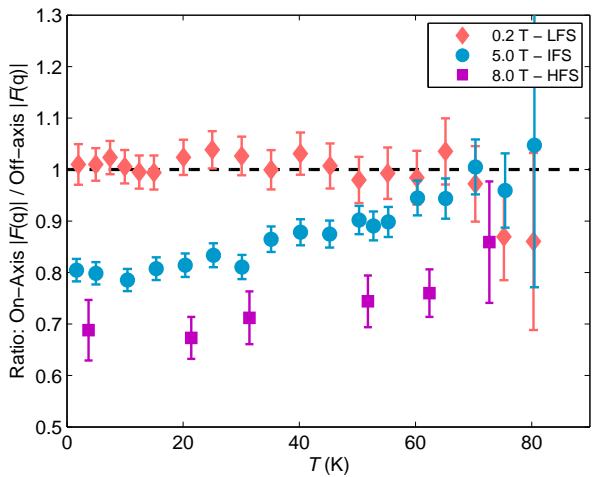}
\caption{(Color online) The temperature-dependence of the form factor ratio taken between the different types of Bragg spots at at various fields. For the fields of 0.2~T and 8.0~T, taken in the LFS and HFS phases respectively, the form factor ratio plotted at each temperature is $\left|F(\textbf{q}\parallel \textbf{a}^{\ast})\right|/\left|F(\textbf{q}\not\parallel \textbf{a}^{\ast})\right|$. For the 5.0~T data taken in the IFS phase, the ratio is $\left|F(\textbf{q}\parallel \textbf{b}^{\ast})\right|/\left|F(\textbf{q}\not\parallel \textbf{b}^{\ast})\right|$.}
\label{fig:T Dep Ratios}
\end{figure}

Despite the ambiguity caused by possible VL disorder, a term in the form of the DW factor cannot explain the emergence of the low $T$ form factor anisotropy between the different types of Bragg spot. Instead, this anisotropy is likely to arise as a consequence of field-induced non-local effects on the VL structure. This is further inferred by an examination of the $T$-dependence of the form factor ratio. Data for selected fields are presented in Fig.~\ref{fig:T Dep Ratios} where, as for Fig.~\ref{fig:form_factor_ratios_2K}, at each $T$ we plot the form factor ratio between the on-axis and off-axis spots. Within the LFS phase, the data at 0.2~T show the ratio to remain at unity over the entire $T$ range, which is as expected for a VL close to the local London regime. At 5.0~T the form factor ratio, which is less than unity at low $T$, smoothly tends towards unity on the approach to $T_{\textrm{c}}$. These results show that increasing $T$ suppresses the low $T$ form factor anisotropy that is induced by nonlocal effects on the VL. The persistence of this behavior to fields within the HFS phase is also tentatively supported by our results obtained at 8.0~T. This indicates that it may be possible to move between a square-like phase with four strong spots and two weaker spots (a VL like that shown in Fig.~\ref{fig:structures}~(c)) towards a hexagonal-like structures with six spots of similar intensity (like that shown in Fig.~\ref{fig:structures}~(a)) just by increasing $T$. We expect future SANS measurements to shed light on the validity of this suggestion.

\subsubsection{Phase Diagram for $\mu_{0}H$~$\parallel$~\textbf{c}}

\begin{figure}
\includegraphics[width=0.5\textwidth]{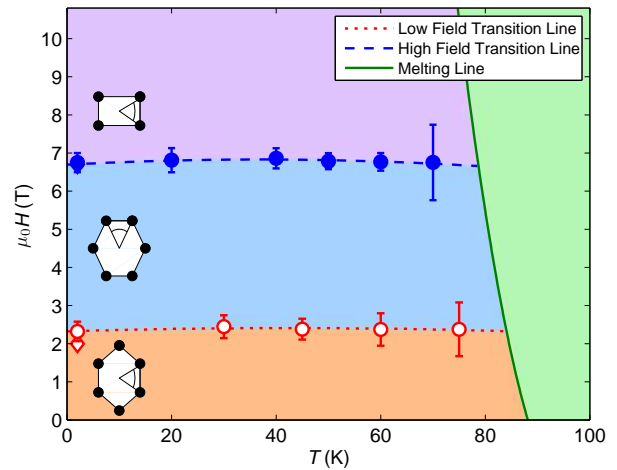}
\caption{(Color online) The ($\mu_{0}H$,$T$) VL structure phase diagram for fields applied parallel to the \textbf{c}-axis. Circle data points were obtained from measurements of the VL form factor, and correspond to the point where each relevant VL structure is measured to occupy 50\% of the sample volume. Diamond datapoints correspond to estimates of the transition points as determined from structural measurements. The dotted and dashed lines are guides for the eye phase boundary lines determined from the form factor data only. The melting line is deduced from data presented in Ref.~\onlinecite{Rou98}.}
\label{fig:Phase_Diagram}
\end{figure}
In this final section, we summarize the main results of our study in the form of the $\mu_{0}H$~$\parallel$~\textbf{c} phase diagram shown in Fig~\ref{fig:Phase_Diagram}. Datapoints were determined from measurements of either the VL structure or the VL form factor. For the LFS - IFS transition, our best estimate of the transition field at 2~K using form factor data is 2.3(2)~T, a field slightly higher than that determined from the structural data alone. For the IFS - HFS transition at 2~K, the transition field determined by both structural and form factor measurements are in good agreement, both being at 6.7(2)~T. This phase diagram shows more clearly how the phase boundary lines between different VL structure phases remain essentially constant in field across the entire $T$ range, a result consistent with that inferred from measurements of the $T$-dependence of the VL structure. While we can not rule out any curvature in either phase boundary line closer to $T_{c}$, the SANS measurements show no evidence for a strong variation.

We also draw comparison between the phase diagram shown in Fig~\ref{fig:Phase_Diagram}, and that obtained from SANS studies on a lightly twinned sample of YBa$_{2}$Cu$_{3}$O$_{7}$.~\citep{Whi08} The at best weakly $T$ dependent phase boundaries observed in the twin-free sample are in contrast to the single phase boundary line observed in the twinned sample, which clearly curves upwards in field at higher $T$.~\citep{Whi08} Furthermore, as the structure phase boundary for the twinned sample is continuous, we suggest that its field and $T$ phase diagram is determined by a complex balance between the intrinsic $\mathcal{F}$ of the high field VL, and the extrinsic $\mathcal{F}$ associated with the $\{110\}$ pinning potential.

\section{SUMMARY AND CONCLUSIONS}
\label{sec:4Summ}
In this study we have reported a small-angle neutron scattering (SANS) study of the VL in a high quality, and twin-free, sample of YBa$_{2}$Cu$_{3}$O$_{7}$. The absence of vortex pinning to twin planes allows the intrinsic VL structure to be imaged, and provides the possibility to make more direct comparisons between theoretical prediction and field- and temperature ($T$)-dependent observations than is otherwise possible in twinned samples.

Our measurements of the VL structure at 2~K show that a field-driven sequence of first-order structure transitions exist in twin-free YBa$_{2}$Cu$_{3}$O$_{7}$. The driving mechanisms behind these transitions can be broadly described as caused by the increasing importance with field of non-local effects on the intervortex interaction. Theoretically, non-local effects are catered for by the development of higher-order correction terms to local theories. These terms play the role of coupling the VL to anisotropies of the host material. One source of anisotropy expected to influence the VL properties is that due to the band structure,~\citep{Kog96,Kog97a,Kog97b,Suz10} with another source arising due to the predominantly $d_{x^{2}-y^{2}}$-wave order parameter anisotropy.~\citep{Aff97,Fra97}

Within the theoretical literature, we find that no model is able to provide a full description of the field-dependent sequence of VL structure transitions that we observe. The simplest theory that seems most appropriate for making predictions of the field-dependent VL structure in twofold symmetric systems like YBa$_{2}$Cu$_{3}$O$_{7}$ is provided by Kogan~\emph{et al}.~\citep{Kog97a,Kog97b} This model accounts for nonlocal effects that arise due to the band structure anisotropy, and not superconducting gap anisotropy. Our close examination of the Kogan theory has revealed some hitherto unreported properties which show that in its current form, the model framework is over-simplified and unable to explain field-dependent observations in \emph{real} twofold systems. Nevertheless, in spite of the noted shortcomings of the model, there are indications within our data to suggest that nonlocal effects that originate from the band structure anisotropy~\citep{Kog97a,Kog97b} may be the cause of our low field transition between orthogonal hexagonal VL structures. Similar arguments can be made to suggest that the high field transition between hexagonal and rhombic VL structures has the same origin. However, this second transition can also be understood as caused by the increasing influence of the predominantly $d$-wave gap anisotropy, whether this tendency is captured by a non-local theory~\citep{Aff97,Fra97} or by detailed microscopic calculations.~\citep{Ich99,Nak02,Suz10} Based on our structural data alone, it is difficult to decide which of these two sources of anisotropy causes the high field square-like structure to appear. The available predictions lead us to expect a high field square-like VL structure with an orientation consistent with that which we observe, regardless of whether band structure or pairing anisotropy is dominant. However, measurements at higher $T$ show each of the structure phase boundary lines to exhibit the same weak $T$-dependence, suggesting that a common mechanism may lie behind both transitions.

In the lowest field VL structure phase, local anisotropic London theory~\citep{Kog81,Cam88,Thi89} appears to provide an adequate description for the VL distortion and form factor. While such a theory is unable to explain the stabilization of a preferred orientation (which \emph{must} be due to weak non-local interactions), it does provide a framework within which we can obtain a measure of the in-plane anisotropy parameter $\gamma_{\lambda}=\lambda_{a}/\lambda_{b}$, either directly from the measure of the distortion of the low field hexagonal structure, or from fitting parameters obtained in an analysis of the field-dependence of the form factor at 2~K.

An important finding is that a regime close to the local limit ends abruptly on moving into the higher field structure phases. While the strongest evidence for high field non-local effects is provided by our measurements of the VL structure, other evidence is found on studying the field- and $T$-dependence of the VL form factor. In particular, the emergence of a form factor anisotropy that becomes increasingly significant at larger fields, and which is quenched with increasing $T$, points towards the decisive role that non-local effects play on the observed VL structure. Furthermore, within the intermediate field phase at 5.0~T, we observe an unusually weak low $T$-dependence of the VL form factor for both types of Bragg spot. An analysis using a simple model that includes a non-local correction suggests that this unusual behavior may arise due to non-local effects induced by the anisotropy of the $d$-wave order parameter.~\citep{Ami98,Ami00} While such a $d$-wave non-local theory~\citep{Ami98,Ami00} may explain the observed weak low $T$-dependence of the VL form factor at 5.0~T, there is poor agreement between the VL structures that we observe, and those predicted within the same theoretical framework.~\citep{Aff97,Fra97,Ami98} This is likely because these models do not consider a vortex core anisotropy that will arise due to anisotropies in either the band structure or the gap function, and which other theoretical studies show can be responsible for field-driven transitions between different VL structures.~\citep{Ich99,Nak02,Suz10} Overall it seems that the precise field- and $T$-dependent VL properties are sensitive to a delicate balance between the field-dependent anisotropies in the system. The high quality data presented in this study should serve to inspire theoretical interest in the precise properties of the VL in High-$T_{\textrm{c}}$ YBa$_{2}$Cu$_{3}$O$_{7}$. Further progress in understanding these results will also benefit from new SANS studies at yet higher magnetic fields.


\begin{center}
\textbf{ACKNOWLEDGEMENTS}
\end{center}
We acknowledge valuable discussions with E.~Blackburn, M.R.~Eskildsen, M.Ichioka, V.G.~Kogan, K. Machida, M. Reibelt, and A. Schilling. Experiments were performed using the D11 instrument at Institut Laue-Langevin, Grenoble, France, the SANS-I and SANS-II instruments at the Swiss spallation neutron source SINQ, Paul Scherrer Institut, Villigen, Switzerland, and the NG3-SANS instrument at the NIST Center for Neutron Research, Gaithersburg, USA. Experiments using NG3-SANS are supported in part by the National Science Foundation under Agreement No. DMR-0454672. We also acknowledge the support of NIST and the U.S. Department of Commerce, in providing the neutron research facilities used in this work. Financial support is acknowledged from the EPSRC of the UK, the Swiss NCCR program MaNEP, the DFG in the consortium FOR538, DanScatt, and from the European Commission under the 6th Framework Programme through the Key Action: Strengthening the European Research Area, Research Infrastructures, Contract No. RII3-CT-2003-505925.


\bibliography{YBCO}

\begin{thebibliography}{10}
\bibitem{Not03}%
  \BibitemOpen
  \bibinfo {note} {In the low-field region, just as in the fourfold case, there
  are two degenerate $\sim$45$^{\circ}$ VLs; the second one is as shown in
  Fig~\ref{fig:kogan_two} but reflected about a short-dashed mirror plane.}%
  \bibAnnoteFile{Stop}{Not03}%
\bibitem{Not04}%
  \BibitemOpen
  \bibinfo {note} {We regard a transition as second-order if the VL structure
  does not have a discontinuity at the transition, with the VL angle as a
  thermodynamic order parameter; as mentioned in the text, the discontinuous
  changes of VL structure that we observe [Fig~\ref{fig:opening_angles}] are
  regarded as first-order.}%
  \bibAnnoteFile{Stop}{Not04}%
\bibitem{Not01}%
  \BibitemOpen
  \bibinfo {note} {Here we use the definition
  $\lambda_{i}\propto\sqrt{m_{i}^{\ast}}$, such that currents along $i$ screen
  against varying fields along $j$ or $k$.}%
  \bibAnnoteFile{Stop}{Not01}%
\end{thebibliography}

\end{document}